\documentclass[12pt]{article}

\usepackage{amsmath,epsf,amssymb,latexsym,cite,xy}
\xyoption{matrix}\xyoption{arrow}

\usepackage[dvips]{epsfig}

\newcommand{\beq}{\begin{equation}}
\newcommand{\eeq}{\end{equation}}
\newcommand{\be}{\begin{equation}}
\newcommand{\ee}{\end{equation}}
\newcommand{\beqa}{\begin{eqnarray}}
\newcommand{\eeqa}{\end{eqnarray}}

\newif\iffigs\figstrue

\DeclareFontFamily{U}{rsf}{}
\DeclareFontShape{U}{rsf}{m}{n}{
   <5> <6> rsfs5 <7> <8> <9> rsfs7 <10-> rsfs10}{}
\DeclareMathAlphabet\Scr{U}{rsf}{m}{n}

\def\pplogo{\vbox{\kern-\headheight\kern -43pt
\halign{##&##\hfil\cr&{
\ppnumber}\cr\rule{0pt}{2.5ex}&\ppdate\cr}
}}
\makeatletter
\def\ps@firstpage{\ps@empty \def\@oddhead{\hss\pplogo}%
   \let\@evenhead\@oddhead 
}
\def\maketitle{\par
  \begingroup
  \def\thefootnote{\fnsymbol{footnote}}
  \def\@makefnmark{\hbox{$^{\@thefnmark}$\hss}}
  \if@twocolumn
  \twocolumn[\@maketitle]
  \else \newpage
  \global\@topnum\z@ \@maketitle \fi\thispagestyle{firstpage}\@thanks
  \endgroup
  \setcounter{footnote}{0}
  \let\maketitle\relax
  \let\@maketitle\relax
  \gdef\@thanks{}\gdef\@author{}\gdef\@title{}\let\thanks\relax}
\makeatother

\def\H{{\mathbb H}}

\def\Large{\large}

\def\half{\frac {1}{2}}

\numberwithin{equation}{section}

\def\thefootnote{\fnsymbol{footnote}}

\def\Im{\,{\rm Im}\, }
\def\Re{\,{\rm Re}\, }

\def\BC{{\mathbb C}}
\def\IC{{\mathbb C}}
\def\IR{{\mathbb R}}
\def\IZ{{\mathbb Z}}
\def\be{\begin{equation}}
\def\ee{\end{equation}}
\def\ba{\begin{eqnarray}}
\def\ea{\end{eqnarray}}

\renewcommand{\H}{{\cal H}}

\newcommand{\p}{\partial}

\newcommand{\Rop}{\mathbb{R}}

\def\bX{{\bar X}}

\def\bra#1{\langle #1 |}

\def\bket#1{|\!|#1\rangle\!\rangle}

\def\Imm{{\rm Im}}
\def\Ree{{\rm Re}}

\begin{document}

\vspace*{-1.5cm}
\thispagestyle{empty}
\begin{flushright}
BRX-TH-581\\
NSF-KITP-06-125\\
hep-th/0702036
\end{flushright}
\vspace*{1.0cm}

\begin{center}
{\Large
{\bf Bulk perturbations of N=2 branes}}
\vspace{1.0cm}

{\large Matthias R.\ Gaberdiel}%
\footnote{{\tt E-mail: gaberdiel@itp.phys.ethz.ch}}

\vspace*{0.2cm}

Institut f{\"u}r Theoretische Physik, ETH Z{\"u}rich\\
CH-8093 Z{\"u}rich, Switzerland\\
\vspace{0.5cm}

{\large {and}} 
\vspace{0.5cm}

{\large Albion Lawrence}%
\footnote{{\tt E-mail: albion@brandeis.edu}}

\vspace*{0.2cm}

Theory Group, Martin Fisher School of Physics, \\
Brandeis University, MS057, PO Box 549110 ,Waltham, MA 02454 USA\\
{\it and}\\
Kavli Institute for Theoretical Physics, University of California\\
Santa Barbara, CA 93106 USA

\vspace*{1.5cm}

{\bf Abstract}

\end{center}

The evolution of supersymmetric A-type D-branes 
under the bulk renormalization group flow between two different $N=2$
minimal models is studied. Using the Landau-Ginzburg description 
we show that a specific set of branes decouples from the infrared
theory, and we make detailed predictions for the behavior of the
remaining branes. The Landau-Ginzburg picture is then checked against
a direct conformal field theory analysis. In particular we construct a
natural index pairing which is preserved by the RG flow, and show that
the branes that decouple have vanishing index with the surviving 
branes.

\newpage
\renewcommand{\theequation}{\arabic{section}.\arabic{equation}}

\renewcommand{\thefootnote}{\arabic{footnote}}
\setcounter{footnote}{0}

\section{Introduction}

Renormalization group flows for perturbed 2d conformal field theories
without boundary have been studied in great detail, as have 2d
conformal field theories on spaces with boundaries, where the 
theory is perturbed by a relevant boundary operator. However, the flow
of boundary conditions under relevant bulk perturbations is poorly
understood, and only specific aspects of
some examples have been studied: see  
\cite{Adams:2001sv,Martinec:2002wg,Minwalla:2003hj,Moore:2004yt,%
Moore:2005wp,Adams:2005rb} for flows to nontrivial IR fixed points, and 
\cite{Ghoshal:1993tm,LeClair:1995uf,Chatterjee:1995be,%
Dorey:1997yg,Dorey:1999cj,Dorey:2004xk,Green:2006ku} for massive
flows. For exactly marginal bulk perturbations, on the other hand, a
general description of the flow of boundary conditions has
recently been found in \cite{Fredenhagen:2006dn}. One may hope that
the techniques that were developed there may also shed light on the 
case with relevant bulk perturbations. 

We would expect this problem to have interesting applications in
condensed matter physics. It should also be important for string
theory. In many cases, renormalization group (RG) flows of worldsheet
conformal field theories (CFTs) provide a qualitatively sensible
description of the dynamics of tachyon condensation in spacetime
(see for example \cite{Freedman:2005wx,Graham:2006gc} for recent
discussions of this point). For closed string tachyons which are not
localized, the dilaton initially evolves towards strong coupling 
\cite{Freedman:2005wx}, indicating that D-branes 
could be an important part of the dynamics of closed string tachyon
condensation.\footnote{However, in the study of collapsing circles
with winding tachyons, the authors of \cite{McGreevy:2005ci} find
evidence that all perturbative strings become highly massive and
perturbation theory remains valid.} 
\smallskip

One particular class of theories where one may hope to be able to
address the flow of boundary conditions under bulk perturbations are 
the 2d unitary minimal models with or without supersymmetry. 
In these models, there exists only a finite set of boundary conditions
which preserve the full (super)conformal symmetry. In the 
simple example of a charge-conjugation modular invariant 
these boundary conditions are in one-to-one correspondence with the
chiral primaries of the minimal model \cite{Cardy:1989ir}. 
As we perturb the theory by a
(supersymmetry-preserving) relevant operator, the bulk theory will
flow to a minimal model with a smaller central charge for which the
number of chiral primaries is smaller.  In particular, the
number of allowed boundary conditions therefore also decreases. It is
then an interesting question to study what happens to the various
boundary conditions of the UV theory under the flow. This is essentially
the same question, in conformal field theory language, as the one
asked in \cite{Martinec:2002wg}.

In this work we shall give an answer for a wide class of RG flows of 
2d $N=2$ supersymmetric minimal models with boundary, perturbed by
relevant bulk F-terms. There are advantages and disadvantages to
studying the flows in this class. The disadvantage is that 
unlike the case of the non-supersymmetric minimal models, the RG fixed
points are separated by an infinite distance as measured by the
Zamolodchikov  metric \cite{Cvetic:1989qv}, so that conformal
perturbation theory is not a useful tool.  The advantage is that the
D-branes which preserve the 
$N=2$ supersymmetry are governed entirely
by the F-terms of the bulk theory, which satisfy non-renormalization
theorems that allow one to understand the RG flow completely.  

In this paper we shall study the so-called A-type branes that have 
a very explicit description in terms of the Landau-Ginzburg (LG)
analysis of \cite{HIV}. This will allow us to make quite detailed
predictions for the behavior of these branes under the bulk
perturbation. We shall find that certain branes flow to
superconformal branes of the IR theory, while others decouple. These 
predictions can be tested against direct conformal field theory
arguments, and we find perfect agreement.
In particular, inspired by the the Landau-Ginzburg analysis,
we identify the RR-charges of the IR theory in terms of the
RR-charges of the UV theory. We then construct an index in the UV
conformal field theory that only takes into account these IR 
RR-charges, and that flows to the standard index of \cite{HIV} in the
infrared. 
It can therefore be used to measure the IR RR-charges of any UV brane;
among other things this allows us to distinguish the branes that
decouple (and have vanishing index), from those that flow to
non-trivial branes in the IR.   
In addition we are able to
study the Affleck-Ludwig $g$-function \cite{Affleck:1991tk} for
the branes which decouple along the flow, and we find that $g$ always
tends to zero for them\footnote{Note that there is no reason for the
`$g$-theorem' -- the statement that $g$ decreases along RG flows 
induced by relevant boundary perturbations -- to hold for {\it bulk}
perturbations.}. The physical interpretion of this result is that the
corresponding branes decay.\footnote{Note that refs.  
\cite{Martinec:2002wg,Moore:2004yt,Moore:2005wp}, in the case
of the decay of nonsupersymmetric orbifolds following \cite{Adams:2001sv},
argue that the kinetic terms for the associated RR charges
vanish as well.} 
\bigskip

The paper is organized as follows.  
In \S2 we review the $N=2$ minimal models, their RG flows, and the
corresponding Landau-Ginzburg (LG) description.
In \S3\ we explain the CFT and LG descriptions of A-type D-branes in
minimal models.  
In \S4\ we study the bulk RG flows from the point of view of
the Landau-Ginzburg theory, and identify which branes decouple in this
framework. We also study a number of examples in detail and compute
the IR limit of the $N=2$ generalization \cite{HIV} of the
Affleck-Ludwig `$g$-function' \cite{Affleck:1991tk}.  In \S5\ we
analyze the problem using CFT methods. In particular, we construct an
index for the D-branes which is preserved along the RG flow. This then
allows us (at least for a large class of examples) to confirm the
above RG predictions. We conclude in \S6\ with some remarks on the
implications for closed string tachyon dynamics, and a discussion of 
possible future directions for research.
There are two appendices where some of the more technical arguments
are described in detail.

\section{Review of minimal models}

We begin by reviewing some basic facts about $N=2$ minimal models
(more precisely, the `A series' of minimal models) and their
Landau-Ginzburg description.

\subsection{The N=2 superconformal field theory}  

The holomorphic $N=2$ superconformal algebra is generated by the
stress tensor $T(z)$, a $U(1)_R$ current $J(z)$, and supersymmetry
currents $G^{\pm}(z)$ with $U(1)_R$ charges $\pm 1$. For $c \leq 3$
there are a discrete set of unitary `minimal models' at central charge 
\be
c = \frac{3k}{k+2} \ , \qquad k=1,2,\ldots \ ,
\ee
containing a finite number of superconformal primaries. The
corresponding (bulk) theories have an A-D-E classification. 
The theories in the `A series' of the $N=2$ minimal models have the 
spectrum 
\be\label{bulk}
\H = \bigoplus_{[l,m,s]} \H_{[l,m,s]} \otimes
\overline{\H}_{[l,-m,-s]} \ ,
\ee
where the sum runs over all equivalence classes of representations of 
the coset algebra 
\be\label{coset}
\frac{ {\rm su}(2)_k \oplus {\rm u}(1)_{2k+4}}{{\rm u}(1)_{4}} 
\ee
that describes the bosonic subalgebra of the 
$N=2$ algebra.\footnote{For a more detailed description of these
standard conventions see for example \cite{Recknagel:1998sb}.} 
The representations of (\ref{coset}) are labelled by the triples  
$(l,m,s)$, where $l=0,1,\ldots, k$; $m=0,1,\ldots, 2k+3$; and
$s=0,1,2,3$. The triples have to satisfy the constraint that $l+m+s$
is even; in addition we have the identification 
$(l,m,s)\simeq (k-l,m+k+2,s+2)$, where both $m$ and $s$ 
are being regarded as periodic variables with period $2k+4$ and $4$,  
respectively. The ground state in $\H_{[l,m,s]}$ 
has conformal dimension $h$ and $U(1)_R$ charge $q$
\be
h = \frac{l(l+2) - m^2}{4(k+2)} + \frac{s^2}{8}\ \hbox{(mod 1)} \ ;
\qquad q = \frac{m}{k+2} - \frac{s}{2}\ \hbox{(mod 1)} \ .
\ee

The coset representations with $s$ even live in the NS sector,
while those with $s$ odd are R-sector representations. In the
following the chiral and anti-chiral primaries will also play an
important role: in the NS sector, the chiral primaries appear in the
sectors $(l,l,0)$ or $(l,-l-2,2)$. The anti-chiral primaries arise for
$(l,-l,0)$ or $(l,l+2,2)$. In the R sector, the ground states (or
primaries) arise in the sectors $(l,l+1,1)$ or $(l,-l-1,-1)$.

\subsection{Landau-Ginzburg description}

The A-series minimal models can be described as the IR fixed
point of a Landau-Ginzburg theory for a single complex scalar
superfield $X$ 
\cite{Zamolodchikov:1986db,Kastor:1988ef,Vafa:1988uu,Martinec:1988zu}:
in particular, the $k$th minimal model is the IR fixed point of the
theory 
\be\label{eq:lgaction}
{\cal L}_k = \int d^2x\, d^2\theta d^2\bar\theta \, \bX   X 
- \int d^2x\, d^2\theta \, W(X) + {\rm h.c.}  
\ , \qquad W(X) = \frac{\lambda_k}{k+2} X^{k+2} \ .
\ee
The chiral primaries of the theory are simply the bottom components of
the superfields $X,X^2,\ldots X^k$ ($X^{k+1}$ is a descendant operator
by the equations of motion), with $X^l$ corresponding to the operator
$(l,l,0) \otimes \overline{(l,-l,0)}$ at the conformal fixed point.
The antichiral primaries are labelled by $\bX^l$ corresponding to
$(l,-l,0)\otimes \overline{(l,l,0)}$.

\subsection{Renormalization group flows of perturbed minimal models}

Next we want to perturb ${\cal L}_k$ by the top component of the
superfield for a relevant chiral primary
\be\label{eq:perturbedlg}
{\cal L}_k \to {\cal L}_k - \int d^2x\, d^2 \theta 
\frac{\lambda_{n}}{n+2} X^{n+2}\ +\  {\rm h.c.} \ ,
\ee
where $n < k-1$.  In terms of the superpotential this simply
corresponds to adding to the superpotential of (\ref{eq:lgaction}) the
term 
\be\label{eq:relevantlgpert}
W(X) = \frac{\lambda_k}{k+2} X^{k+2} + 
\frac{\lambda_{n}}{n+2} X^{n+2} \ . 
\ee
Standard non-renormalization theorems indicate that
no additional superpotential terms are induced.  Assuming that the
kinetic term is in the right universality class, the most relevant
coupling $\lambda_{n}$ should dominate, and the theory will flow 
at low energies to the $n$th minimal model.\footnote{The superpotential
is not renormalized -- however, the fields $X$ undergo
wavefunction renormalization.  By rescaling $X$ appropriately, the
RG flow can be seen as taking $|\lambda_{n}|\to\infty$ and
$|\lambda_k|\to 0$ as one flows to the 
IR; this is what one might expect from
the fact that $\lambda_{n}$ is the most relevant coupling in
(\ref{eq:relevantlgpert}).} 

\begin{figure}[ht]
\centering
\epsfig{file=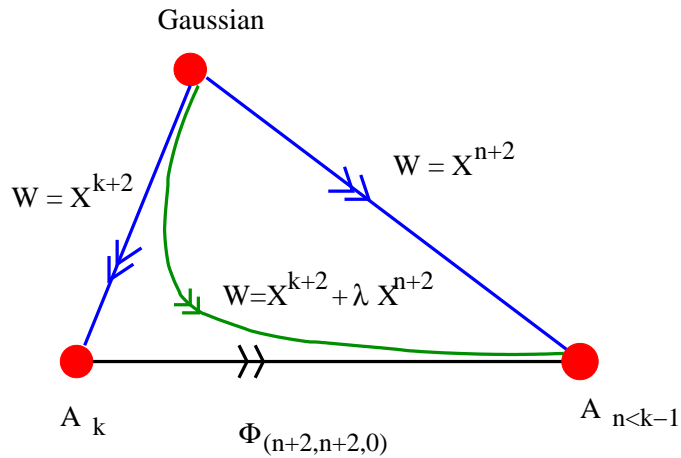,height=6cm}
\parbox{5.5in}{\caption{A map of RG flows.  The top corner of the
triangle denotes the free $c=3$ $N=2$ superconformal theory of a
free chiral  multiplet $X$. Deforming this theory by the
superpotential $W=X^{k+2}$ ($X^{n+2}$) induces a flow to the $k$th
($n$th) minimal model $A_k$ ($A_n$).  Upon a deformation by the
relevant chiral primary labelled by $(L,M,S) = (n+2,n+2,0)$, $A_k$
flows to $A_n$. Finally, the superpotential 
$W = X^{k+2} + \lambda X^{n+2}$ induces a flow from the free theory  
which asymptotes to the  the flow between the free theory and
$A_k$ in the UV, and to the flow between $A_k$ and $A_n$ in the IR.}}
\end{figure}

This flow approximates the flow from the $k$th minimal model in the  
UV to the $n$th minimal model in the IR, under the perturbation by 
$(n+2,n+2,0)\otimes \overline{(n+2,-(n+2),0)}$. Strictly speaking,
the perturbation (\ref{eq:relevantlgpert}) induces a flow from the
free theory in the UV  to the $n$th minimal model. However, if
$\lambda_{n}$ is small in the UV, the RG flow should begin close to
the flow between the free theory and the $k$th minimal model; as
$\lambda_{n}$ gets large, the flow starts to run close to the line
between the $k$th and $n$th minimal models, as shown in figure~1.
In this way we can say that the RG flow of (\ref{eq:perturbedlg})
approximates the RG flow we are interested in, at least for large
$\lambda_{n}$. Note that this basic idea, (and the RG diagram in
figure~1) is essentially identical in spirit to the use of gauged linear
sigma models for describing the decay of tachyons in
non-supersymmetric orbifolds
\cite{Martinec:2002wg,Moore:2004yt,Moore:2005wp}. (Indeed, in those
references, the RG flow of these orbifolds is discussed using
the mirror of the gauge linear sigma model, which is a 2-field
Landau-Ginzburg theory \cite{Moore:2005wp,Hori:2000kt,Vafa:2001ra}.) 

In trying to understand the behavior of D-branes under RG flow, we
will therefore assume (as do \cite{Martinec:2002wg,Moore:2004yt,Moore:2005wp})
that studying D-branes in the LG theory
(\ref{eq:relevantlgpert}) as $\lambda_{n}\to\infty$ will be a good
guide to the behavior of D-branes in the flow between the $N=2$ minimal
models $A_k$ and $A_{n}$. The essential reasoning is that the
D-branes we will be studying are controlled entirely by the F-terms,
and the RG flow of these F-terms drives $\lambda_{n}$ to infinity such
that it dominates over $\lambda_k$ in the infrared. Furthermore, we
shall also be able to check at least some of our conclusions  against
direct conformal field theory arguments, which gives us confidence in
our reasoning based on Landau-Ginzburg theory.

\section{D-branes in minimal models}

In the following we shall study the branes (or boundary conditions)
that preserve the full $N=2$ superconformal algebra at the
boundary. Such boundary conditions fall into two clases, 
`A-type' and `B-type' branes, that are distinguished by 
their gluing condition for the $U(1)_R$ current \cite{Ooguri:1996ck}. 
In this paper we will focus on the `A-type' boundary conditions, as
they have a very simple description in the Landau-Ginzburg theory.  

\subsection{A-type branes in conformal field theory}

A-type branes are characterised by the gluing conditions  
\begin{eqnarray}
(L_n - \bar{L}_{-n}) |\!| B\rangle\!\rangle & = & 0 \nonumber \\
(J_n + \bar{J}_{-n}) |\!| B\rangle\!\rangle & = & 0 \label{Atype} \\
(G^\pm_{r} + i \eta \bar{G}^\pm_{-r}) |\!|B\rangle\!\rangle & = & 0 
\ . \nonumber 
\end{eqnarray}
Here $\eta=\pm$ corresponds to the two different spin structures; in
the following we shall concentrate on one choice of $\eta$, say 
$\eta=+$.

In conformal field theory these branes are described by their 
boundary states that can be constructed following Cardy 
\cite{Cardy:1989ir}. In each sector of (\ref{bulk}) there is an
Ishibashi state $|l,m,s\rangle\!\rangle$ satisfying (\ref{Atype}).
The consistent boundary states are then labelled by the
representations of the coset algebra, {\it i.e.}\ by triples $[L,M,S]$
satisfying the same range (and identifications) as above. Explicitly,
the boundary state corresponding to $[L,M,S]$ is given as 
\be\label{eq:cardystate}
|\!| L,M,S\rangle\!\rangle = 
(2k+4)^{-1/4} \sum_{[l,m,s]} 
\frac{S_{Ll}}{\sqrt{S_{0l}}} \, 
e^{i\pi \frac{Mm}{k+2}}\, e^{-i\pi \frac{Ss}{2}} \, 
|l,m,s\rangle\!\rangle \ ,
\ee
where the sum runs over all equivalence classes of coset
representations, and $S_{Ll}$ are the S-matrix elements of SU(2) at
level $k$
\be
S_{Ll} = \sqrt{\frac{2}{k+2}}\, 
\sin\left(\pi\, \frac{(L+1)\, (l+1) }{k+2} \right)\, .
\ee
The two different choices of $\eta=\pm$ correspond now to the
different choices $S$ odd or $S$ even. Since we are only interested in
one of the two choices we require $S$ to be odd from now
on. 
\smallskip

\noindent The bulk theory (\ref{bulk}) has RR ground states in the
sectors $(l,l+1,1)\otimes \overline{(l,-l-1,-1)}$.
For the following it is convenient to normalize these RR ground states
as 
\be\label{RRstates}
e_l = \frac{1}{\sqrt{S_{0l}}} \, 
(l,l+1,1)\otimes \overline{(l,-l-1,-1)} \ .
\ee
With this convention the brane denoted by $|\!|L,M,S\rangle\!\rangle$
has charge $Q_l(L,M,S)$ with respect to (\ref{RRstates}), where 
\be\label{charges}
Q_l(L,M,S) = 
\left(\frac{2}{(k+2)^3}\right)^{\tfrac{1}{4}} \, 
\sin\left( \pi \frac{(L+1) (l+1)}{k+2} \right) 
e^{i \pi \frac{M (l+1)}{k+2}} \,  e^{-i\pi \frac{S}{2}}  \ . 
\ee

\subsection{Landau-Ginzburg description of the A-type branes}

As was explained in \cite{HIV}, we can identify the A-type $N=2$
D-branes with directed lines in the complex $X$-plane 
(where the orientiation corresponds to the sign of the RR charge of
the brane).
The requirement that the brane preserves A-type
supersymmetry implies that $\Im W(X)$ is constant along these
curves. However, not all such lines are of interest since 
in general the corresponding brane will have infinite RR charge (and
tension). In fact, the $m$'th RR charge of the corresponding D-brane   
\be
g_{m,k} =  \langle X^m |B\rangle\!\rangle
\ee
is a topological quantity \cite{Ooguri:1996ck}. (Here we have denoted
a basis of the RR ground states by $X^m$, reflecting the fact that 
the RR ground states which couple to an A-type boundary state are in
one-to-one correspondence with the $(c,c)$ chiral ring of the
B-twisted theory, generated by the field $X$.) For A-type branes 
the topological field theory localizes on constant maps into the
$X$-plane \cite{HIV,Ooguri:1996ck}, and this can be used to show 
\cite{HIV} that 
\be\label{eq:normalizedg}
g_{m,k} =N(m,k) \int_{\gamma} dx\, x^m \, e^{-W(x)} \ , 
\ee
where $\gamma$ denotes the curve and $N(m,k)$ is a 
normalization factor that turns out to be 
\be\label{eq:gcoefficient}
N(m,k) = \frac{1}{\Gamma\left(\frac{m+1}{k+2}\right)}
\sqrt{\frac{k+2}{2 \sin\left(\frac{\pi(m+1)}{k+2}\right)}}\ . 
\ee
Since this calculates the RR charges of the branes, we should require
that $g_{m,k}$ be finite. This implies that $\Re W$ must be bounded
from below along $\gamma$. In particular, $\gamma$ must pass through a 
critical point of $W$, {\it i.e.} through $x =x_0$ with 
$W'(x_0) =0$. For a homogenous superpotential of the form
(\ref{eq:lgaction}) all critical points coincide at the origin. Since
$W(0)=0$, the relevant lines have all  $\Im W=0$. The different curves
must therefore run along the lines 
\be\label{lines}
z = t\, e^{\frac{2 \pi i r}{k+2}} \ , \qquad t\in\Rop^+ \ , \qquad 
r= 0, 1 \ldots, k+1 \ . 
\ee
We will call the brane $(r_1,r_2)$ the brane that corresponds to the
line for which the orientation is towards the origin along $r_1$ and 
towards infinity along $r_2$.
This curve corresponds to the A-type brane (\ref{eq:cardystate}) with
\be\label{ident}
L+1 = |r_2-r_1| \ , \qquad M = r_1 + r_2 \ , \qquad 
S = {\rm sign} (r_2-r_1) \ . 
\ee
Note that the pair with the opposite orientation corresponds to the
other possible value of $S$; it thus describes the anti-brane. The
identification (\ref{ident}) was checked in \cite{HIV} by showing that
the RR-charges $Q_l(L,M,S)$ agree with the corresponding charges 
calculated from (\ref{eq:normalizedg}).

The geometric interpretation, together with the orientation, makes it
clear that we can generate all branes from the $L=0$ branes.  That is,
consider the branes $(r,r+1)$ and $(r+1,r+2)$.  Both branes have a
branch which lies along the $z = t e^{2\pi i (r+1)/(k+2)}$ line, but
with opposite orientation, as seen on the left hand side of figure~2.
We expect that in the sum of these branes, a tachyon develops along
this line, and that the two branes  merge to form the brane $(r,r+2)$,
as shown on the right hand side of figure~2. This picture is borne out
by an explicit analysis in conformal field theory 
\cite{Fredenhagen:2001nc,Maldacena:2001ky}. In particular, one easily
checks that the RR-charges $Q_l(L,M,S)$ are preserved along this
boundary RG flow.  

\begin{figure}[htb]
\centering
\epsfig{file=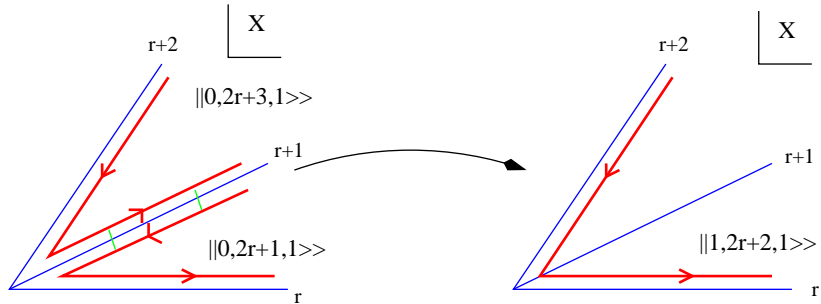,height=4cm}
\parbox{5.5in}{\caption{Building all Cardy states from $L=0$ branes.
The blue lines labeled $r,r+1,r+2$ are the lines for which 
$W \in \IR_+$. The red lines, forming wedges between  them, are
offset slightly so as to make them better visible.
On the left they denote two adjacent $L=0$ branes, with the sign of the charge denoted
by their orientation. The green lines between these branes are open
string tachyons which stretch between branes which lie on the same
line but have opposite orientation.  The $L=1$ brane on the right is
the endpoint of open string tachyon condensation.}} 
\end{figure}

\subsection{Perturbing the branes}

Now we want to consider a perturbation of the form 
(\ref{eq:relevantlgpert}), as a tool for understanding the
behavior of D-branes under RG flow from the minimal model $A_k$ in the
UV to the minimal model $A_{n}$ in the IR.  Our basic philosophy, as
discussed in \S2.3\ above, is that this behavior can be understood by
studying D-branes  in the $|\lambda_{n}|\to\infty$ limit.  
At $\lambda_{n} = 0$, the branes can be mapped to the  
minimal model $A_k$ describing our UV fixed point, using 
the techniques of \cite{HIV}.

\subsubsection{Some general comments}

The claim that we can follow the branes simply by following the
contours of constant $\Imm\ W$ as $|\lambda_{n}| \to\infty$ may
require some further explanation. First, under RG flow the bulk
D-terms also get renormalized; however, as shown in \cite{HIV}, these
do not affect the configurations of the A-type branes. Secondly, one
may wonder whether any boundary operators are induced along the flow,
as in  \cite{Fredenhagen:2006dn, aelunpub, Green:2006ku}.  
The non-renormalization theorem for the superpotential also holds for
half-superspace integrals on the boundary \cite{Hori:2000ic}. 
Marginal and relevant boundary D-terms, on the other hand, will have
the form of gauge field couplings, $A(X) \p_0 X$; however, in the
present case where we  are not including Chan-Paton factors, such
terms can be re-written as a gauge-trivial $B$-field coupling in the
bulk, which is known not to change the equation $\Imm\ W = 0$ for the
A-type branes. Furthermore, A-type branes emanating from isolated
zeros are rigid, so we do not expect any marginal or relevant
operators of the form  $T(X) \p_1 X$ to be induced on the boundary.  

As a further check, if the non-renormalization of the F-terms holds
directly in the perturbed minimal models, only D-terms will be
induced. However, all of the nontrivial boundary D-terms will 
be irrelevant.  The reason is that they will be of the form 
$\int d^2\theta \Psi$, where $\Psi$ is a superconformal primary
field. For minimal models all nontrivial superconformal
primaries other than the identity have positive dimension, so the
above term in the action will always have dimension greater than $1$.
Therefore, in a scheme in which only marginal and relevant couplings
are induced by RG flow to the IR, we should not need to consider any
boundary terms. This could change if the bulk flow induces irrelevant 
terms to become relevant, but since the number of superconformal
primaries decreases and the spectrum of conformal dimensions remains
gapped with only the identity at dimension zero, 
we believe this will not happen.

\subsubsection{Brane decay}

When $\lambda_{n}$ is small, we are describing a small perturbation of the
UV theory. The asymptotic behavior (for large $x$) of the
corresponding curves should not be modified. This implies that each
curve must continue to approach two of the UV-lines (\ref{lines}) as
we go to large $x$. (We shall call the lines (\ref{lines}) for the
$k$th minimal model the `UV-lines' in the following.) We can thus
label each of these curves by a pair $(r_1,r_2)$ as above, and for
$\lambda_{n}$ small, these lines can again be identified with the
branes of the (perturbed) conformal field theory via (\ref{ident}).  

For each of these curves, $\Re W(x)$ increases as $x$ goes to 
infinity. This implies that the derivative of the real part of 
$W$ must be zero somewhere along the line, and hence that the line
must still go through a critical point of the superpotential
$W'(x_0)=0$.

The description above implies that not all combinations of $(r_1,r_2)$
will continue to be described by contours of constant $\Imm\ W$ 
\cite{HIV}. To see this let us assume that the critical point $x_0$ 
is isolated. $W(x)$ can be approximated as 
$W(x) = W(x_0) + \half W''(x_0) (x- x_0)^2 + \cdots$ in the 
neighborhood of $x_0$. There are then two branches with 
$\Im W(x) = \Im W(x_0)$ for which $\Re W(x) \geq \Re W(x_0)$ and these
describe the brane of interest. Therefore, the asymptotic 
behavior of the brane is already uniquely determined by the choice of 
an isolated critical point. For example, if all critical points are
isolated, we will only have $k+1$ possible curves, whereas the
original unperturbed superpotential had ${k+2 \choose 2}$ different
curves.  

One may thus wonder what happens to the branes of the UV theory that
do not satisfy this constraint. At least in the examples we shall
study  we shall be able to show that any brane of the UV
theory can be written as a superposition of branes (in the sense of
figure~2) that continue to exist. It is then very plausible to believe
that upon switching on the perturbation, a generic UV brane (that
cannot continue to exist) will simply decay into a superposition of  
branes that survive the perturbation, perhaps (for small $\lambda$)
still fused together at large $|x|$ by open string tachyons.  
\smallskip

In any case, as we switch on the perturbation (\ref{eq:perturbedlg})
the Landau-Ginzburg description will allow us to analyze the subset of
branes of the original $k$th minimal model theory that are still
compatible with this constraint. In particular, we can study their
behavior as we increase $\lambda_{n}$. This will give us a
predicition for how these branes of the $k$th minimal model behave
under the RG flow to the $n$th minimal model. This is what we shall
be studying in the following. We shall also see that our findings have
a natural interpretation in conformal field theory.

\subsection{$g$-function}

At the conformal fixed point, the $0$'th RR charge is related to the 
Affleck-Ludwig $g$-function by a phase, $g=|g_{0,k}|$.  In boundary
conformal field theory, $g$ describes the coupling of the boundary
state to the identity, $g = \langle 0|\!|B\rangle\!\rangle$, where
$\bra{0}$ is the $SL(2,\BC)$-invariant vacuum in the closed string
sector \cite{Affleck:1991tk}. This measures 
a regularized dimension of the Hilbert space of open strings beginning
and ending on this brane \cite{Affleck:1991tk,Harvey:1999gq}. In
string theory compactifications for which the minimal model is a
factor of the `internal' part of the CFT (as opposed to the spacetime
factor  described by a sigma model on ${\mathbb{R}}^4$), the 4d
graviton will contain the identity operator acting on the minimal
model factor, and the $g$-function will compute the tension of the
brane in four dimensions \cite{Harvey:1999gq}.  

The $g$-function decreases along {\em boundary} renormalization group
flows \cite{Affleck:1991tk}, and for such flows it evolves via a
gradient flow \cite{Friedan:2003yc}. For relevant {\it bulk}
perturbations, however, there is no reason to believe such a theorem
will hold, and we need to be specific about how we might even define $g$ in
such a theory.

On the other hand, since the $g$-function is directly related to the
tension of the corresponding brane in string theory, it is important
to understand how it behaves under the above RG flows, at least in the
IR limit when $|\lambda_{n}|\rightarrow \infty$. When RG flow is
interpreted as time evolution (see
\cite{Freedman:2005wx,Graham:2006gc} for reviews and further
references), this will be important data for understanding the
behavior of D-branes under time evolution. 

\subsubsection{Definition of the $g$-function}

A physically motivated definition, which matches the definition at the
bulk conformal point, is the overlap of the boundary state with the
properly normalized NS vacuum of the Landau-Ginzburg theory.
Similar definitions have appeared in 
\cite{LeClair:1995uf,Chatterjee:1995be,Dorey:1997yg,Dorey:1999cj}
(and an alternative definition proposed in \cite{Dorey:2004xk}.)
This NS vacuum is well-defined for
any value of the coupling constants $\lambda_{k}$ and $\lambda_{n}$,
and it becomes the $SL(2,\IC)$-invariant vacuum at the fixed points of
(\ref{eq:perturbedlg}).  In a string  theory setting, one may make the
perturbed theory physical by coupling it to the spacetime coordinates 
as in \cite{Freedman:2005wx}. If the time evolution is not too
rapid, we believe that for the NS ground state 
of the coupled theory, used to define the 4d graviton,
the wavefunction in the LG directions will be 
well-approximated by the wavefunction for the ground state of the LG
theory. 

Unfortunately, this overlap is quite complicated in general, and one
may therefore be tempted to continue using 
(\ref{eq:normalizedg}) for $m=0$, which would be much easier to
compute. However, it does not have a clear physical interpretation
away from the conformal point. First of all, one should not expect
that the tension of the brane ({\it i.e.}\ the coupling to the
$SL(2,\BC)$-invariant vacuum in the closed string sector) is directly
related to a RR-charge away from the conformal point. Furthermore, we
cannot expect that (\ref{eq:normalizedg}) continues to make sense for
$\lambda_{n}\rightarrow \infty$ since in this limit some of the 
RR ground states have disappeared altogether.\footnote{This can also
be seen directly in the derivation of \cite{HIV} where a rescaling of
the superpotential is implicit. If $\lambda_{n}$ is finite this
washes away the lower order perturbation described by $X^{n}$.}

In order to calculate the $g$-function we therefore need to evaluate
the overlap with the ground state. The precise ground state
wavefunction is difficult to determine, but we can make some
qualitative statements. For example, consider the perturbation 
(\ref{eq:perturbedlg}) for large $\lambda_{n}$.  In the NS sector,
the bosonic potential $|W'(X)|^2 = |X^{k+1} + \lambda X^{n+1}|^2$ has
its minima at the zeros of $W'(X)$. Apart from the $(n+1)$st-order
zero at the origin, there are $k-n$ isolated zeros at the solutions of
$X_c^{k-n} + \lambda_{n} = 0$.  Fluctuations about these vacua have
masses of order $|W''(X_c)| \propto |\lambda_{n}|^{\frac{k}{k-n}}$,
which becomes infinite as $|\lambda_{n}|\to\infty$.\footnote{We should
be careful as the {\it physical}\ mass will depend on the coefficient
of the kinetic term, which is renormalized.  In fact we are assuming
that the family of theories with canonical kinetic term and $W$ as in
(\ref{eq:relevantlgpert}), indexed by  rescalings of the norm of
$\lambda_{n}$, are a good model of RG flow for the gross properties of
the D-branes that we are interested in.} In the NS sector there are no
fermion zero modes.  The wavefunctions in field space of the various
momentum modes in this massive theory are well-approximated by the
wavefunctions for the corresponding simple harmonic oscillator, and
their ground state energies will grow with $\lambda_{n}$.  Therefore,
we expect all of the low-energy states to be concentrated at the
origin in $X$-space and die off rapidly away from the origin. The
boundary state for any D-branes whose support is far from the origin
of $X$-space will thus have vanishingly small overlap with the ground
state. For such branes one therefore expects that the $g$-function
goes to zero in the limit $|\lambda_{n}|\to\infty$.  This suggests
that the graviton ceases to couple to these branes.  Furthermore all
indications are that the RR fields sourced by the branes themselves
decouple from the theory
\cite{Martinec:2002wg,Moore:2004yt,Moore:2005wp}.  

\subsubsection{Physical interpretation}

However we do not think that the fact that $g\to 0$ for these branes
indicates that they become light; in fact, this would be a potential
disaster for the theory, as it would mean that any string background
which is a possible endpoint of closed string tachyon condensation
would have massless nonperturbative objects. Rather, the above
discussion indicates that they completely decouple from the
excitations described by the infrared conformal field theory. They do
not interact with such fields, and we believe they cannot be excited
by such fields. For all {\it practical} purposes, from the point of
view of the observer described by the infrared conformal field theory,
the D-branes associated to the critical points moving to infinity have
decayed away.  

As in \cite{Freedman:2005wx}, 
this RG analysis will not capture the full time-dependent
spacetime process of closed string tachyon decay in the presence of a
D-brane, but it should be an important first step in that direction.
We close this section by noting to what degree our results are
(or are not) similar to the discussions in
\cite{Adams:2005rb,Green:2006ku}. 
We are studying branes that remain attached to massive vacua
in the IR, rather than branes which are localized at the tachyon
condensate.  In the case of \cite{Adams:2005rb}, this would
be like branes localized at the `dust vacua'.  An analogous
setup would be one in which the Riemann surface in \cite{Adams:2005rb}
is split into two Riemann surfaces, and the observer lives in
one component while a D1-brane wraps a 1-cycle in the other component.
It seems consistent that such a brane will be seen as having decoupled
from the observer. The case of \cite{Green:2006ku}\ would
be most analogous to a perturbation of the form $W = X^2$ in our 
LG picture, for which all of the vacua are massive; our techniques
would focus on branes attached to the massive vacua, and would
be similar to D0-branes localized at the zeros of the tachyon in 
\cite{Green:2006ku}.

\section{Branes in perturbed theories: LG analysis}

Now that we have set up the study of supersymmetric boundary
conditions in general Landau-Ginzburg models, we can analyze how the  
above branes behave under the bulk RG flow between minimal models. 
As we have seen, the A-type supersymmetric branes are determined
entirely by the superpotential. We will thus assume that the questions
we are asking can be addressed by studying the family
(\ref{eq:relevantlgpert}) as one rescales $\lambda_{n}$. 
It is easy to track the various branes in this family by following the 
level lines of $\Im W$ emanating from the solutions to $W'(x) = 0$.

In this section we will do this for the case that $n+2$ is a factor of
$k+2$. In \S5\ we will find a clean conformal field theory
interpretation of our results in the case that $(k+2)/(n+2)$ is odd.

\subsection{Perturbation by $W = X^{(k+2)/d}$}

We now want to study the case where we perturb the
$k$'th minimal model by the relevant bulk field $X^{n+2}$, where $n+2$
is a factor of $k+2$. The superpotential is thus of the form
\be\label{pert1}
W(x) =  \tfrac{1}{k+2} \, x^{k+2} + \lambda \, \tfrac{d}{k+2} \,
x^{\frac{k+2}{d}} \ ,
\ee
where $d$ is a factor of $k+2$. 
Part of the reason that this
class of examples is particularly easy is that the perturbed theory
still has the non-trivial rotation symmetry
\be\label{eq:flav}
	\rho: x \to e^{\frac{2\pi i d}{k+2}}\, x \ . 
\ee
Furthermore, if $\lambda$ is real, the conjugation transformation  
\be\label{eq:flip}
	C: x \to e^{\frac{2\pi i d}{k+2}}\, x^*
\ee
is a symmetry of the equations of motion $\Im W = 0$. 
We will find that these symmetries cut our workload considerably.

In the following we shall focus on the case $\lambda \in\IR_+$. The
structure of the UV and IR branes are as follows.  In the limit
$\lambda \to 0$ the branes are described by the lines (\ref{lines}), 
while in the limit $\lambda\to\infty$, the branes are described by the
IR lines 
\be\label{eq:irlines}
x = t e^{\frac{2\pi i d r}{k+2}}\ ;\qquad t \in \IR_+\ ,\qquad 
r=0,1,\ldots, n+1 \ . 
\ee
We will also call the wedges between adjacent UV (IR) lines $r,r+1$
the $(r+1)$st `UV(IR) wedges'; the branes bounding each UV(IR) wedge
are thus the $L=0$ branes in the UV(IR) theory (see figure~3 for an
example).  Note that the transformation $\rho$ in (\ref{eq:flav}) maps
the $(r+1)$st IR wedge into the $(r+2)$nd IR wedge.

Let us further denote by $B$ the line which bisects the first IR
wedge, {\it i.e.} the line with angle $\tfrac{\pi d}{k+2}$. By
construction, $W(x)$ is real along $B$ (as well as along the real
axis and the other IR lines). 
The $B$-line will lie on the
$d/2$nd UV line when $d$ is even, and it will bisect the $(d+1)/2$nd
UV wedge when $d$ is odd. Furthermore, the transformation $C$
(\ref{eq:flip}) will flip the $x$-plane around $B$, and maps the first
IR wedge back into itself. 

\begin{figure}[htb]
\centering
\epsfig{file=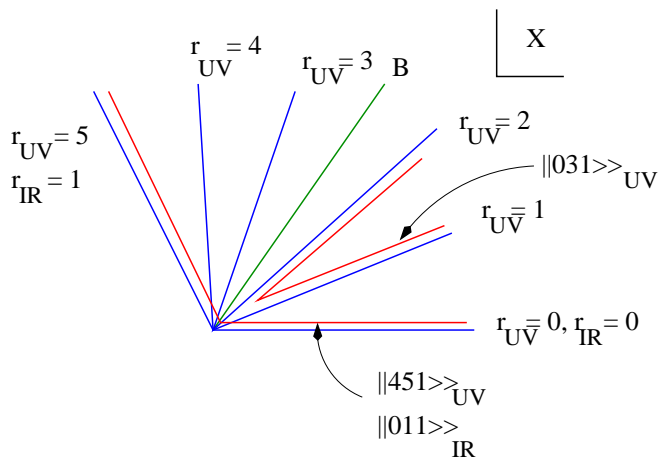,height=6cm}
\parbox{5.5in}{\caption{UV and IR lines, and some sample D-branes, for
the case $d=5$.}} 
\end{figure}

It is now immediate that the IR branes
form a subset of the UV branes. In particular some of the curves of
the UV theory stay completely fixed upon taking 
$\lambda\rightarrow \infty$. From the point of view of the
Landau-Ginzburg description, the corresponding branes therefore do
not change shape under the  RG flow (see figure~3 for an
example). Translated into conformal field theory language, this then
suggests that we have the flow   
\be\label{eq:irtoir}
|\!| pd-1,d(p+2q),\pm 1\rangle\!\rangle_{\rm UV} \quad 
\Longrightarrow \quad 
|\!| p-1,(p+2q),\pm 1\rangle\!\rangle_{\rm IR} \ , \qquad
p =1,\ldots,n+1 \ . 
\ee
We will also find some support for this conjecture in the conformal
field theory analysis of \S5. 
\smallskip

On the other hand, most of the UV branes do not appear in the IR
theory since there are more UV lines than IR lines. We wish to understand
the fate of the other UV branes as $\lambda\to\infty$. 
For $\lambda\neq 0$, the critical points of $W$ are at $x=0$ (with 
multiplicity $n+1$), as well as at the $(k-n)=\tfrac{(k+2)(d-1)}{d}$ 
points
\be\label{eq:critpoints}
x_l = \lambda^{\frac{d}{(k+2)(d-1)}} \, 
e^{\pi i \frac{d (2l+1)}{(k+2)(d-1)}} \ ,
\qquad l=0,\ldots,\frac{(k+2)(d-1)}{d} - 1 \ . 
\ee
Note that $(d-1)$ of these lie in the interior of each adjacent IR wedge.
In particular, $x_{l=0,\ldots,d-2}$ lie in the first IR wedge
between $r = 0,1$ in (\ref{eq:irlines}). For future reference we also
note that the value of $W$ at these critical points is 
\be\label{eq:Watcrit}
W(x_l) = \tfrac{d-1}{k+2} \lambda^{\frac{d}{d-1}} \, 
e^{\pi i \frac{(2l+1)}{d-1}}  \ . 
\ee

Let us focus on the zeros lying within a single IR wedge -- by the
symmetry $\rho$ 
(\ref{eq:flav}), the story will be identical for the
remaining IR wedges. The details of which branes decouple and
why depend somewhat on whether $d$ is odd or even.
We will describe the general pattern here, and discuss details
for the case $d$ odd and even separately below.

Let $\lambda$ have a small imaginary part. In this case, the UV and 
IR lines deform slightly.  The line $B$ defined by $\Imm\ W = 0$, which
lies in the interior of the IR wedge, then bisects a UV wedge. This is  
because upon so deforming $\lambda$, no critical point $x_l$ will
satisfy $\Imm\ W(x_l) = 0$; in particular, $B$ therefore does not go
through a critical point, and since $\Ree\ W$ decreases near the
origin along $B$, it must continue to do so as $|x|$ becomes large along $B$. 
The $B$ line therefore cannot asymptote to an UV line along
which $\Ree\ W$ increases as $|x|$ becomes large. 

In \S4.2, \S4.3, we shall find that all other UV wedges not bisected by
$B$ contain exactly one critical point, 
and (with the aforementioned deformation of $\lambda$) no critical
points lie on a UV line. The branes that emanate from the non-trivial 
critical points cannot cross $B$ or any of the IR lines (along which
we also have that $\Imm\ W = 0$) since by the above assumption the
value of $\Imm\ W$ at the critical points does not vanish.
Furthermore, on all of the UV lines in
this region, $W(x)$ is real and positive (for $|x|\rightarrow
\infty$), and the phase of $W$ near each line increases (decreases) as
the phase of $x$ increases (decreases). Thus $\Imm\ W$  is slightly
positive above each UV line, and slightly negative just below each UV
line, as illustrated in figure~4. Using these two facts, the results
of Appendix A show that every 
UV $L=0$ brane that asymptotes to the boundary of a wedge containing a
critical point, can be generated 
(by open string tachyon condensation) from branes that decouple.

\begin{figure}[htb]
\centering
\epsfig{file=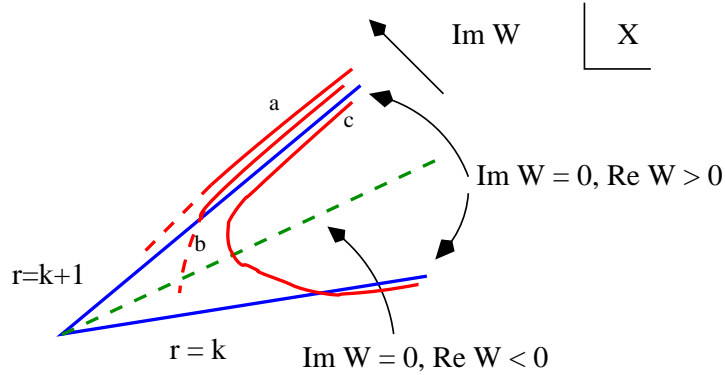,height=5cm}
\parbox{5.5in}{\caption{Three contours with constant $\Im W$ close to
zero.  All such contours asymptote to UV lines. Close to the UV lines,
$\Imm\ W$ increases as shown. Branes asymptoting from a
counterclockwise direction to a UV line have $\Im W > 0$; branes
asymptoting from a clockwise direction have $\Im W < 0$.  In this
picture $\Im W_a > \Im W_b > \Im W_c$.}}
\end{figure}

The one $L=0$ brane for which this argument does not apply (and which
will in fact not decouple) is the one asymptoting to the 
boundary of the UV wedge bisected by $B$. Let us call this brane 
$S$. As we have argued above, the branes associated to the non-trivial
critical points decay along the flow. By the same arguments as given
below for $d$ odd, one can then show that the UV brane $S$ must flow
to the IR $L=0$ brane associated to the IR wedge in question. For
example, for the case depicted in figure~3, all of the branes 
$|\!|0,2p+1,1\rangle\!\rangle_{\rm UV}$ in the first IR wedge will
decouple in the IR except for $p = 2$. On the other hand, the $L=0$
brane $|\!|0,5,1\rangle\!\rangle_{\rm UV}$ flows to the $L=0$ brane of
the IR wedge, {\it i.e.}\ to the brane 
$|\!|011\rangle\!\rangle_{\rm IR}$.  

{}For $d$ odd, we will find strong evidence in conformal field theory
that this picture holds, as we will see in \S5. We do not have such an
argument for $d$ even, but we suspect that this is just a technical
matter.

\subsection{The case of odd $d$}

{}From the LG perspective the case where $d$ is odd is simpler 
since the $B$ line already bisects a UV wedge even for real $\lambda$. 
The decay picture we have just sketched is then
unambiguously defined. [For $d$ even and $\lambda \in \IR_+$, the line
$B$ lies on top of the $d/2$nd UV line.  If we resolve the situation
by making $\lambda$ slightly complex,  the UV wedge associated 
to $S$ depends on the sign of $\Imm\ \lambda$.]

\noindent Returning to the case $d$ odd, the critical points $x_l$
have the phases  
\be\label{eq:criticalphases}
\theta = \frac{\pi i }{k+2} \theta_l \ , \qquad \hbox{with} \qquad 
\theta_l =  (2l+1) \frac{d}{d-1} \ .
\ee
These points always lie between adjacent UV lines, since
\be
0 < \frac{d}{d-1} < 2 < \frac{3d}{d-1} < 4 < \frac{5d}{d-1} \cdots \ .
\ee
The first UV wedge that does not contain a critical point is the
$(d+1)/2$nd one, since
\be
d-3 < \theta_{\frac{d-3}{2}}=\tfrac{(d-2)\, d}{d-1} < d-1 < d+1 < 
\theta_{\frac{d-1}{2}} = \tfrac{d^2}{d-1} < \ldots \ . 
\ee
Using the transformation $C$ (\ref{eq:flip})  it follows that all
other UV wedges in the 
first  IR wedge have one and only one critical point in their
interior. In fact, this had to be the case since there are 
$(d-1)$ critical points in the first IR wedge, and $d$ UV wedges, and
because the angle between consecutive zeros is greater than the angle
between adjacent UV lines (as can be seen from
(\ref{eq:criticalphases}) and (\ref{lines})). Finally, by
the symmetry $\rho$ in eq.\ (\ref{eq:flav}), the same story repeats
itself in each IR wedge. The discussion in \S4.1, together with
the proof in Appendix A, now shows that all of the $L=0$
branes, except for the $L=0$ brane bisected by $B$ and its images
under $\rho$, decouple from the IR conformal field theory. 

Actually, we can be more explicit and identify concretely which branes
emanate from these zeros. Let us concentrate on the critical points
between the real axis and $B$; these are the zeros $x_l$ with
$l=0,\ldots,\tfrac{d-3}{2}=u$. The symmetries $\rho$ and $C$ will tell
us what the curves through the other zeros look like. Note that it
follows from (\ref{eq:Watcrit}) that  
\be\label{Wvals}
\Imm\ W(x_l) = \Imm\ W(x_{u-l}) \ , \qquad l=0,\ldots,
\lfloor \tfrac{u}{2} \rfloor \ .
\ee

We can prove (see appendix~B) and have verified by numerical
computation for $d \leq 13$ the following geometric picture for the
branes emanating from $x_l$. The curve for $l=0$ asymptotes to the
zero'th and first UV line, {\it i.e.} it  
corresponds to the brane $\bket{0,1,\pm 1}_{\rm UV}$ in conformal
field theory. The curve through the critical point
for $l=u=\tfrac{d-3}{2}$ asymptotes in one direction to the 
$\tfrac{d-1}{2}$th UV line.  In the opposite direction it goes to the
critical point $l=0$; because of (\ref{Wvals}) the imaginary parts
of $W$ agree and $\Re W(x_u)< \Re W(x_0)$. From there the brane may
take either branch of the brane through $x_0$, {\it i.e.} either
branch of $\bket{011}_{UV}$. If we give $\lambda$ a small positive 
imaginary part, $\Imm\ W(x_u) < \Imm\ W(x_0)$.  The contours
now cannot cross; furthermore, given the discussion in the previous
section and appendix~B, 
the line from $x_u$ must asymptote to either the 0th or 1st
UV line slightly counterclockwise from the brane emanating from
$x_0$. Therefore, it must asymptote to the real axis, and hence
correspond to the brane $\bket{u,u+1,\pm 1}_{\rm UV}$ in conformal
field theory.

\begin{figure}[htb]
\centering
\epsfig{file=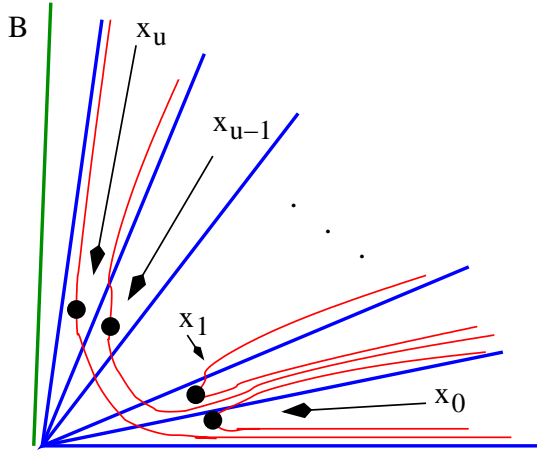,height=6cm}
\parbox{5.5in}{\caption{A picture of the decoupling branes for 
$d$ odd when $\Imm\ \lambda > 0$.}}
\end{figure}

Continuing in this fashion, we have found that 
the curve through $l=1$ asymptotes to the first and second UV line, 
{\it i.e.} it corresponds to $\bket{0,3,\pm 1}_{\rm UV}$ in 
conformal field theory. The line through $l=u-1$ on the other hand
asymptotes in one direction to the $\tfrac{d-3}{2}$nd UV line,
and joins the $l=1$ critical point in the other; as before, if we give
$\lambda$ a small positive imaginary part, it will continue along the
first UV line, and hence correspond to 
$\bket{\tfrac{d-7}{2},\tfrac{d-1}{2},\pm 1}_{\rm UV}$ in conformal
field theory.
The curves that go through these critical points thus correspond to
the UV branes 
\begin{eqnarray}
& \bket{0,2l+1,\pm 1}_{\rm UV} 
\qquad & l=0,\ldots, \lfloor \tfrac{u}{2} \rfloor  \nonumber \\
& \bket{2l-u,u+1,\pm 1}_{\rm UV} 
\qquad & l=\lfloor \tfrac{u}{2} \rfloor +1, \ldots, u \ ,
\label{decoupbr} 
\end{eqnarray}
where as before $u=\tfrac{d-3}{2}$. As $\lambda\to \infty$, all of
these curves move away from the origin in $x$-space. Since 
$g\to 0$ in this limit, they completely decouple from the IR theory
and can be treated as having decayed away. The same is true for the
branes between $B$ and the first IR line, as can be shown using the 
transformation $C$ in (\ref{eq:flip}).

{}From these decoupling branes (\ref{decoupbr}) we can construct,
using open string tachyon condensation, all of the $L=0$ branes   
$|\!|0,2p + 1,\pm1 \rangle\!\rangle_{\rm UV}$ with 
$p=0,1,\ldots,\tfrac{d-3}{2}$, {\it i.e.} all the $L=0$ branes between
the real axis and $B$. Similarly, from the images of
(\ref{decoupbr}) under $C$ we obtain the $L=0$ branes between $B$ and
the first IR line: $|\!|0,2p + 1,\pm1 \rangle\!\rangle_{\rm UV}$ with  
$p=\tfrac{d+1}{2},\ldots, d-1$. Assuming that the open string tachyon
condensation does not significantly change the 
RG behavior of the component branes, we can thus conclude that all
such $L=0$ branes decouple. As we shall see this will be in agreement
with the conformal field theory index calculation of \S5. 

On the other hand, there is one UV $L=0$ brane in the first IR wedge
for which this argument does not apply, namely 
$|\!|0,d,1\rangle\!\rangle_{\rm UV}$ which asymptotes to the two 
neighboring UV lines that are separated by $B$. 
To determine what this brane flows to we observe that we can write the
UV brane $\bket{d-1,d,1}_{\rm UV}$ asymptoting to the first 2 IR lines
as the sum\footnote{Alternatively we could write it as the sum
$\bket{d-1,d,1}_{\rm UV} = \bket{u,u+1,1}_{\rm UV} 
+ \bket{0,d,1}_{\rm UV} + \bket{u,3u+5,1}_{\rm UV}$ for which we could
also apply the same argument.}
\be\label{eq:irconstruction}
\bket{d-1,d,1}_{\rm UV} = 
\bket{0,d,1}_{\rm UV} 
+ \sum_{p=0}^{\frac{d-3}{2}} \bket{0,2p+1,1}_{\rm UV} 
+ \sum_{p=\frac{d+1}{2}}^{d-1}\bket{0,2p+1,1}_{\rm UV}  \ .
\ee

\begin{figure}[htb]
\centering
\epsfig{file=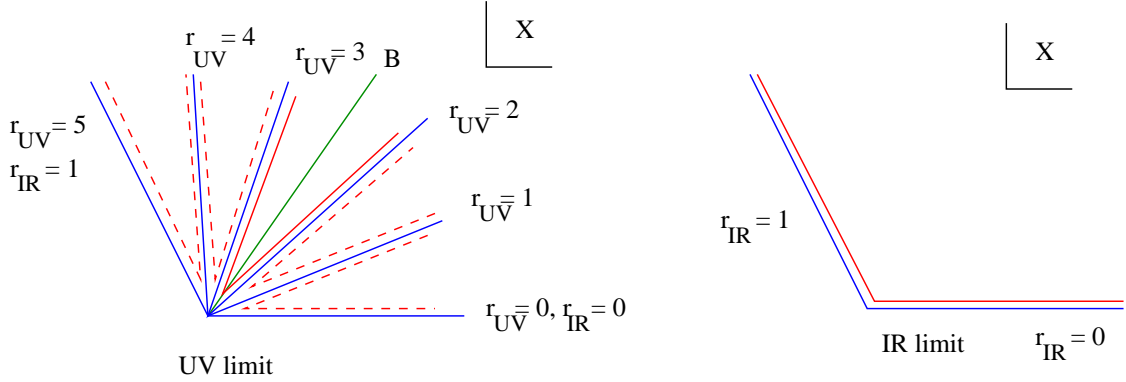,height=5cm}
\parbox{5.5in}{\caption{The flow of D-branes for the case $d=5$. On the
left-hand-side, the dashed lines correspond to $L = 0$ branes which
decouple, while the $L = 0$ UV brane bisected by $B$ flows to the IR
brane $|\!|011\rangle\!\rangle_{\rm IR}$.}}
\end{figure}

\noindent Because of the above results, the branes in the two sums on
the right hand side of (\ref{eq:irconstruction}) all decouple. 
The brane on the left hand side flows as in
(\ref{eq:irtoir}). If we assume that open string tachyon
condensation does not modify the behavior under the bulk flow, we
conclude that $\bket{0,d,1}_{\rm UV}$ flows as\footnote{For any
$\lambda\neq 0$, there is no single curve that corresponds to the
brane $\bket{0,d,1}_{\rm UV}$ any more --- see the discussion in
\S3.3. As is argued there, this must mean that upon switching on
$\lambda$ the $\bket{0,d,1}_{\rm UV}$ brane must decay into the linear
combination defined by (\ref{eq:irconstruction}), perhaps glued
together by open string tachyons when $\lambda$ is still
small. Furthermore, at the endpoint of the RG flow the UV branes which
do not survive the flow must decouple. We believe this means that the
open string tachyons gluing them to the remaining IR brane must become
massive at large $\lambda$; in 2d QFT language the corresponding
operators have thus become irrelevant.} 
\be\label{eq:lzeroflow}
\bket{0,d,1}_{\rm UV} 
\Longrightarrow 
\bket{0,1,1}_{\rm IR} 
\ .
\ee
More generally,
\be\label{eq:middlewedgeflow}
\bket{0,d(2r + 1),1}_{\rm UV} \Longrightarrow 
\bket{0,(2r + 1),1}_{\rm IR}\ ,\qquad r \in \IZ   \ , 
\ee
where we have used the symmetry $\rho$ to obtain the result for the
cases with $r\neq 0$. In section \S5\ we will reproduce this result
using conformal field theory arguments.

\subsubsection{Example: $d=5$}

To illustrate these ideas, let us consider the explicit example with
$d=5$. In this case $u=1$, and there are two UV wedges between $B$ and
the real axis, and two corresponding critical points, $x_0$ and $x_1$.
{}Figure~7 shows a numerical plot of the 
curves emanating from these two
critical points, when $\lambda$ has phase $e^{i\pi/10}$ ({\it i.e.}\ a
small positive imaginary part).  The brane emanating 
from $x_0$ corresponds to the deformation of the Cardy state
$\bket{0,1,1}_{\rm UV}$, while the brane emanating from $x_1$ 
corresponds to $\bket{1,2,1}_{\rm UV}$.  
Note that we can write both  
$L=0$ branes by subtracting the first brane from the second and
condensing the corresponding open string tachyon: 
$\bket{0,3,1}_{\rm UV} = \bket{1,2,1}_{\rm UV} 
- \bket{0,1,1}_{\rm UV}$.    

\begin{figure}[htb]
\centering
\epsfig{file=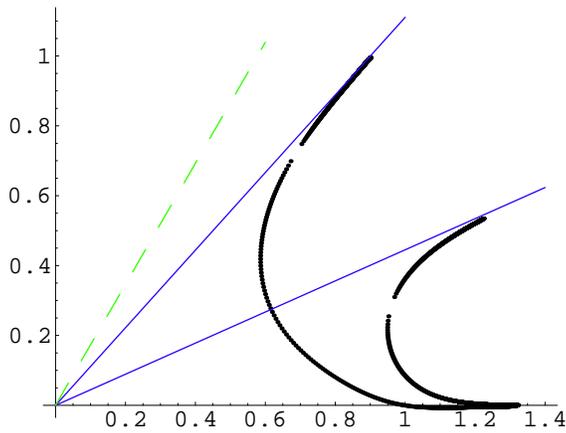,height=6cm}
\parbox{5.5in}{\caption{The decoupling branes in the case $d=5$
between the real axis 
and $B$, where $\lambda$ is given a small imaginary part.  The dashed
green line corresponds to $B$, the horizontal axis is the real axis,
and the vertical axis is the imaginary axis.  The other two straight
lines are the first and second UV lines $r=1,2$.  The curve between
the first UV line and the real axis is the deformation of
$\bket{0,1,1}_{\rm UV}$, and the curve between the real axis 
and the second UV line is the deformation of $\bket{1,2,1}_{\rm UV}$.
The critical points $x_{0,1}$ lie in the breaks of the curves.}}
\end{figure}

We have also studied numerically the curves between $B$ and the first
IR line. When $\lambda$ has a small positive imaginary part, the
transformation $C$ is no longer a symmetry of the equations of motion
-- rather, it takes $\lambda \to \lambda^*$.  For $\Imm\ \lambda > 0$,
the transformation $C$ maps these branes to the branes for the theory
with coupling $\lambda^*$ (for which $\Imm\ \lambda^* < 0$), between
the real axis and $B$.   
Indeed, we find that the the corresponding branes
correspond to $\bket{0,7,1}_{\rm UV}$ and $\bket{0,9,1}_{\rm UV}$.
Therefore,  as discussed above, all of the $L=0$ branes 
between the real axis and the first IR line, except for
$\bket{0,5,1}_{\rm UV}$, 
are generated.  The UV brane $\bket{4,5,1}_{\rm UV}$  
coincides with the IR brane $\bket{0,1,1}_{\rm IR}$ as in
(\ref{eq:irtoir}).  Following (\ref{eq:irconstruction}), 
the UV brane $\bket{0,5,1}_{\rm UV}$ can be written in terms of the
decoupling branes as: 
\begin{eqnarray}
	\bket{0,5,1}_{\rm UV} & = & 
   \bket{4,5,1}_{\rm UV} - \bket{1,2,1}_{\rm UV} - 
   \bket{0,7,1}_{\rm UV} - \bket{0,9,1}_{\rm UV} \nonumber\\
& = & \bket{4,5,1}_{\rm UV} - \bket{0,1,1}_{\rm UV} 
    - \bket{0,3,1}_{\rm UV} - \bket{0,7,1}_{\rm UV} - 
    \bket{0,9,1}_{\rm UV} \ . \nonumber 
\end{eqnarray}
All of the $L=0$ and $L=1$ UV branes on the right hand side decouple,
while we have
$\bket{4,5,1}_{\rm UV} \Longrightarrow\ \bket{0,1,1}_{\rm IR}$,
and hence deduce that 
$\bket{0,5,1}_{\rm UV} \Longrightarrow \bket{0,1,1}_{\rm IR}$ as
well.

\subsection{Even $d$}

The case of $d$ even is quite similar.  However, we do not yet have a
clear CFT picture of this case, so we will be quite brief. 

When $\lambda \in \IR$, the line $B$ and the critical point
$x_{\frac{d}{2}-1}$ lie on the $d/2$nd UV line.  This is a degenerate
situation -- the brane emanating from this critical point lies along
$B$, and upon reaching the origin will continue along either the real
axis or the first IR line.  The degeneracy can be split by giving
$\lambda$ a small imaginary part.  If $\lambda$ has a small positive
imaginary part, the line $B$ defined by $\Imm\ W = 0$ shifts clockwise
and bisects the $d/2$nd UV wedge. On the other hand, if we give
$\lambda$ a small imaginary negative part, $B$ bisects the
$(\frac{d}{2} + 1)$st UV wedge.
In either case, the critical point $x_{\frac{d}{2}-1}$ moves off the 
$d/2$nd UV line, and then it follows from the same arguments as at the
beginning of section \S4.2 that all UV wedges, except for the UV wedge
bisected by $B$, contain precisely one critical point. 

At this point we can appeal to the proof in appendix~A to make the
following claim. For $\Imm\ \lambda$ small and positive,
all of the branes $\bket{0,2p+1,\pm 1}_{\rm UV}$ decouple except for
the brane $\bket{0,d-1,\pm 1}_{\rm UV}$. The UV brane 
$\bket{d-1,d,\pm 1}_{\rm UV}$ asymptotes to the real axis and the
first IR line,  and is expected to flow to $\bket{0,1,1}_{\rm IR}$.
Since this UV brane is a sum of all of the UV branes between the real
axis and the first IR line, we also expect that 
$\bket{0,d-1,1}_{\rm UV} \Longrightarrow \bket{0,1,1}_{\rm IR}$. 

If $\lambda$ has a small negative imaginary part, all of the $L=0$
branes except for $\bket{0,d+1,1}_{\rm UV}$ are expected to decouple,
and we expect that 
$\bket{0,d+1,\pm 1}_{\rm UV} \Longrightarrow \bket{0,1,1}_{\rm IR}$.

\section{D-branes in perturbed minimal models}

As discussed in \S2.3, the LG flow of the action (\ref{eq:perturbedlg})
approximates the flow between the minimal models 
we are interested in.  We therefore expect that our results about the
behavior of the branes under the LG flow apply directly to the flow
between these two conformal field theories. In this section we will
check our conjectures for the  branes discussed in 
\S4.2\ ($d$ odd) directly using conformal field theory methods.

The superpotential (\ref{pert1}) should describe the flow between the
$k$'th minimal model and the $n$'th minimal model that is induced
by the relevant bulk operator from the NS-NS sector
\be\label{ca}
(n+2,n+2,0) \otimes \overline{(n+2,-(n+2),0)} \ , 
\ee
as well as its complex conjugate, the (ac) field 
\be\label{ac}
(n+2,-(n+2)),0) \otimes \overline{(n+2,n+2,0)}  \ . 
\ee
As we have argued above, the LG analysis suggests that we have the
flow (\ref{eq:irtoir}). We shall now attempt to give some support for
this statement from a conformal field theory point of view. We will do
this by computing the conserved RR charges of various D-branes, 
and by constructing a supersymmetric `index' in the UV theory which
flows to the index of \cite{HIV} in the IR theory.  

First we observe that the UV branes that appear on the left-hand-side 
of (\ref{eq:irtoir}), {\it i.e.} the special branes
\be\label{special}
|\!| pd-1,d(p+2q),\pm 1\rangle\!\rangle_{\rm UV}
\ee
have the property that their RR charges (\ref{charges}) satisfy
(recall that $n+2=\tfrac{k+2}{d}$)
\be
Q_{l+(n+2)} (pd-1,d(p+2q),\pm 1) = Q_l(pd-1,d(p+2q),\pm 1) \ , \qquad
l=0,\ldots,k-(n+2)
\ee
and
\be
Q_{r(n+2)-1}(pd-1,d(p+2q),\pm 1) = 0 \ , \qquad
r=1,\ldots,d-1 \ . 
\ee
Thus they only couple to $n+1$ different linear combinations of RR
charges. Furthermore, we have that 
\be
Q_l(pd-1,d(p+2q),\pm 1)_{\rm UV} = {\cal N} \, 
Q_l(p-1,p+2q,\pm 1)_{\rm IR}\ ,  \qquad l=0,\ldots,n \ , 
\ee
where ${\cal N}$ is a constant (independent of $l$), and the charges
on the right hand side refer to the IR theory. This is obviously in 
perfect agreement with the proposed flow (\ref{eq:irtoir}).

There is yet another point of view from which this is very natural.
In the original UV theory the different RR ground states $e_l$ 
(\ref{RRstates}) all have degenerate conformal weight equal to the
minimal value $h=\bar{h}=\tfrac{c}{24}$. As we consider the
perturbation by  
the NS-NS fields (\ref{ca}) and (\ref{ac}), the eigenstates of
lowest conformal weight (that will correspond to the RR ground states
in the IR) will generically be linear combinations of RR ground states
that are related to one another by successive fusions with (\ref{ca})
and (\ref{ac}). Now under successive fusion with the (ca) and (ac)
fields (\ref{ca}) and (\ref{ac}), the RR ground states 
corresponding to  $l+(n+2), l+2(n+2), \ldots, l+(d-1)(n+2)$ 
for $l=0,\ldots,n+1$ get mixed together. Generically, such a mixing
will lift the degeneracy in conformal weight. 
One should thus expect that only one linear combination of these RR
ground states will continue to have minimal conformal weight, while
the conformal weight of the other eigenstates will be bigger. The
eigenstate with minimal conformal weight is then the only state that
can become the RR ground state of the IR theory.\footnote{In fact, the
linear combination corresponding to $l= r (n+2) - 1$ seems to become
massive in the IR.} This is in very nice agreement with the above.  

\subsection{Definition of the index}

We now want to make a prediction for what happens to the other branes,
in particular the UV branes with $L=0$. The above analysis suggests 
an identification of the RR ground states of the IR theory
with specific linear combinations of RR ground states of the UV
theory. With this idea in mind, we should now be able to determine the
IR RR charges of any UV brane; in particular, this can be done using
the index (that calculates an appropriate inner product of the RR
ground states). We do this by studying analogs of the index pairing of
\cite{HIV}.

\noindent In the UV theory, the index is defined by \cite{HIV} (see also
\cite{Brunner:2003zm}):
\be\label{stanindex}
{\cal I}\Bigl(|\!| L_1,M_1,S_1\rangle\!\rangle, |\!|
L_2,M_2,S_2\rangle\!\rangle \Bigr) = (-1)^{\frac{S_1-S_2}{2}} \, 
\left\{
\begin{array}{cl}
N_{L_1 L_2}{}^{\Delta M} & \hbox{if $0\leq \Delta M \leq k$} \\
- N_{L_1 L_2}{}^{-\Delta M - 2} \ & 
\hbox{if $-k-2 \leq \Delta M \leq -2$}  \\
0 & \hbox{if $\Delta M=-1,k+1$}\ . 
\end{array}
\right.
\ee
Here $N_{L_1 L_2}{}^{l}$ is the su$(2)_k$ fusion matrix, 
$\Delta M=M_2-M_1$, and $S_1$ and $S_2$ are both odd. 
This index is not
symmetric, but it is rather obtained from a symmetric inner product
upon applying the spectral flow operator to one of the two branes;
indeed if we denote by $\langle\cdot,\cdot\rangle$ the symmetric inner
product (that is obtained from the usual orthogonal inner product on
the RR ground states) then we have  
\be
{\cal I}\Bigl(|\!| L_1,M_1,S_1\rangle\!\rangle, |\!|
L_2,M_2,S_2\rangle\!\rangle \Bigr) =
\Bigl\langle |\!| L_1,M_1,S_1\rangle\!\rangle, 
|\!|L_2,M_2+1,S_2+1\rangle\!\rangle \Bigr\rangle \ .
\ee
Incidentally, this is the reason for why the index is not symmetric,
but instead satisfies
\be
{\cal I}\Bigl(|\!| L_1,M_1,S_1\rangle\!\rangle, |\!|
L_2,M_2,S_2\rangle\!\rangle \Bigr) =
{\cal I}\Bigl(|\!| L_2,M_2+2,S_2+2\rangle\!\rangle, |\!|
L_1,M_1,S_1\rangle\!\rangle \Bigr) \ . 
\ee
\smallskip

In the application we have in mind, we want to calculate the index
that is relevant to the IR theory, but evaluate it in the UV
theory. One may therefore guess that the relevant spectral flow
operator that enters the index should be the one that is appropriate
for the IR theory, rather than the UV theory. At least for $d$ odd
this is well-defined; we therefore propose that the index we should
use is 
\begin{eqnarray}
\widehat{\cal I}^{(d)}\Bigl(|\!| L_1,M_1,S_1\rangle\!\rangle, |\!|
L_2,M_2,S_2\rangle\!\rangle \Bigr) & = & 
\Bigl\langle |\!| L_1,M_1,S_1\rangle\!\rangle, 
|\!|L_2,M_2+d,S_2+d\rangle\!\rangle \Bigr\rangle   \\
& = & 
{\cal I}\Bigl(|\!| L_1,M_1,S_1\rangle\!\rangle, 
|\!| L_2,M_2+d-1,S_2+d-1\rangle\!\rangle \Bigr) \ . \nonumber
\end{eqnarray}
The index is thus given by the same formula as (\ref{stanindex}),
except that now $\Delta M$ is replaced by 
$\hat{\Delta}M=M_2+d-1-M_1$. 

The index between the special UV branes (\ref{special}) 
and the $L=0$ branes is
\be
\widehat{\cal I}^{(d)}\Bigl(
|\!| pd-1,d(p+2q),\pm 1\rangle\!\rangle, |\!|
0,M,1\rangle\!\rangle \Bigr) = 0 \qquad \hbox{unless $M=d(2r+1)$} \ . 
\ee
Note that this conclusion is actually independent of the order in
which we consider the two branes.  If $M = d(2r + 1)$, we find 
\be
\widehat{\cal I}^{(d)}\Bigl(
|\!| pd-1,d(p+2q),1\rangle\!\rangle, |\!|
0,d(2r+1),1\rangle\!\rangle \Bigr) = 
\epsilon \left(
\delta_{r,p+q-1} - \delta_{r,q-1} \right) \ , 
\ee
where $\epsilon$ is an irrelevant sign. This should be matched by the
index in the IR, for which we have 
\be
{\cal I}_{\rm IR} \Bigl(|\!| p-1,(p+2q),\pm 1\rangle\!\rangle,
|\!| 0 , 2r+1,1\rangle\!\rangle \Bigr) = 
\left(\delta_{r,p+q-1} - \delta_{r,q-1} \right) \ . 
\ee
Thus this suggests the flow
\be
|\!|0,d(2r+1), 1\rangle\!\rangle_{\rm UV} \quad 
\Longrightarrow \quad 
|\!| 0,(2r+1), 1\rangle\!\rangle_{\rm IR}  \ ,
\ee
while all other UV branes with $L=0$ decouple. This rule is in perfect
accord with the results of \S4.2, 
in particular with eq.\ (\ref{eq:middlewedgeflow}). 

This arguments bears some relation to the discussion in
\cite{Moore:2004yt} 
(see also \cite{BR}). In that work the authors construct an index 
pairing, representing the intersection matrix for fractional branes in
a non-supersymmetric orbifold with tachyons; and an intersection
matrix for the D-branes of the theory described by the condensation of
twisted sector tachyons.  They then construct a transformation which
embeds the latter as a block-submatrix of the former; the other block
represents the intersection matrix of the `Coulomb branch branes'
which decouple from the infrared CFT and which bear a strong
resemblance to the decoupling branes in our work (indeed, they 
are mirror to A-type branes in a Landau-Ginzburg theory). In this
section, however, we have provided a {\it construction} of a pairing
in the UV theory directly from a worldsheet point of view, using the
structure of the $N=2$ superconformal algebra. This pairing flows to
the index in the IR, and we have used it to correctly {\it deduce}
which branes of the UV theory will decouple in the IR.   

\section{Conclusions}

In this paper we have determined the behavior of the $N=2$ A-type
branes under a bulk flow from the $k$th minimal model to the $n$th
minimal model
(where $d=\tfrac{k+2}{n+2}$ is an integer). In particular, we
have given good evidence, combining the LG analysis with conformal
field theory arguments, that some of the superconformal branes of the
UV decouple, while others flow to the superconformal A-type branes of
the IR theory. For $d$ odd and real coupling constant $\lambda$, the
flow behavior could be very explicitly described --- see in particular
(\ref{eq:middlewedgeflow}). Using successive open string tachyon
condensation (as for example used in \S4.2) this rule then determines
the fate of every superconformal UV brane under the flow. As far as we
are aware, this is the first time that the behavior of D-branes under
a relevant bulk flow could be determined in such detail.

There are a number of natural directions in which to extend this
work. First of all, there are a few technical points that
deserve further exploration: for example, it would be interesting to
find a CFT argument for $d$ even; 
it would be nice to work out the analysis for the case when $n+2$ is
not a factor of $k+2$ --- probably at least the LG arguments of 
\S4.1\ will generalize fairly straightforwardly to the general situation; 
it would also be good to apply the
same analysis to the D- or E-models (for which everything should go
through). More generally, it would be interesting to study the
fate of the `B-type' branes in this class of RG flows. The CFT
arguments should be quite similar, but the Landau-Ginzburg description
of B-type branes is rather different since they correspond to matrix
factorizations \cite{Kont:unpub,Kapustin:2002bi,Brunner:2003dc} about
which much has been learned in recent years. For the example of the
A-models that has concerned us here, the results of
\cite{Herbst:2004zm} should in particular be useful. Another
interesting exercise would be to study flows in other models in which
we have some control, such as the $N=2$ coset models in
\cite{Bourdeau:1991uu}, which have a Landau-Ginzburg description.  

While the $N=2$ RG flows travel an infinite distance with respect to
the Zamolodchikov metric, the $N=0$ minimal models with 
$c = 1 -\delta$, $\delta \ll 1$ contain both bulk and boundary flows
to neighboring fixed points which are treatable within conformal
perturbation theory \cite{Zamolodchikov:1986db}. The fate of D-branes 
in these models under relevant {\it boundary} flows has been studied 
in \cite{Recknagel:2000ri}, and a generalization of the `absorption of
boundary spin' principle of \cite{Affleck:1990by}\ has been given for
spin-current couplings in more general coset models in
\cite{Fredenhagen:2003xf}. 
It would be interesting to study the fate of boundary conditions under
bulk flows using such techniques. In particular, in a Lagrangian
formulation we would wish to compute the beta functions for coupled
bulk and boundary flows, as in
\cite{Fredenhagen:2006dn,aelunpub,Green:2006ku}, and understand how 
the D-branes evolve. 

Finally, we would like to understand better the implications of this 
work for time-dependent string theory backgrounds with D-branes.

\section*{Acknowledgements}

We would like to thank the KITP where this project was started (as a
result of the `Stochastic Geometry and Field Theory' and `String
Phenomenology' workshops). We would also like to thank Allan Adams,
Paul Aspinwall, Ilka Brunner, Patrick Dorey, Stefan Fredenhagen,
Daniel Green, David Kutasov, Howard Schnitzer, and Eva Silverstein for
helpful conversations and correspondence.  This research was supported
in part by the National Science Foundation under Grant
No. PHY99-07949. M.R.G.\ is partially supported by the Swiss National
Science Foundation and the Marie Curie network `Constituents,
Fundamental Forces and Symmetries of the Universe'
(MRTN-CT-2004-005104). A.L.\ was supported in part by NSF  
grant PHY-0331516, by DOE grant DE-FG02-92ER40706, and by an
Outstanding Junior Investigator award.

\appendix
\renewcommand{\theequation}{\thesection.\arabic{equation}}

\section{Generating all of the decoupling branes}

In this appendix we want to show that the decoupling branes generate
all the $L=0$ charges, except for the $L=0$ brane that is bisected by
$B$. In order for this to make sense we want to make sure that $B$
does not lie on top of a UV line, and that the values of $\Imm\ W$ at
the different critical points are distinct.  For $d$ even or odd, this
is achieved by giving $\lambda$ a small imaginary part. 

\noindent We begin by proving the following statement:
\medskip

\noindent {\bf Claim 1.} For every UV line between two adjacent IR
lines there is at least one decoupling brane ({\it i.e.} a brane whose
corresponding line emanates from a non-trivial critical point) that 
asymptotes to the given UV line.

\vskip .3cm

\noindent{\bf Proof:}
Let us denote the number of UV lines that lie between $B$ and the IR
line asymptoting to the real axis by $m+1$. These lines bound $m$ UV
wedges that lie entirely between $B$ and the IR line, and each of
these wegdes contains precisely one critical point of $W$. (This
follows from the above assumption.) We will label the critical  points
by $x_1,\ldots, x_m$, ordered such that 
$\Im W(x_1) < \Im W(x_2) < \cdots < \Im W(x_m)$.  
We claim that the non-intersecting level lines
through $x_1,\ldots x_m$ will asymptote to at least $m+1$ UV lines. 
The argument will be identical for the branes between $B$ and the 
first IR line.  The images of these branes under $\rho$ will finally
complete the picture. 

Let us label the UV lines in an anticlockwise
order by $L_1,\ldots,L_r$. A given configuration of level
lines is described as follows. For each $L_i$ let $\Delta_i$ be the
ordered set of lines that asymptote to $L_i$ where the order
corresponds to the anti-clockwise order in which the lines approach
$L_i$. We may denote the lines by the critical points through which
they pass; then the order of lines in $\Delta_i$ is the same as the
increasing order of the points $x_j$, and for large $|x|$ is the
increasing order of the corresponding contours in a counterclockwise
direction (see figure~4). We may thus formally write $\Delta_j$ as the
product   
$$
\Delta_j = x_{s_1} x_{s_2} \cdots x_{s_n} \qquad 
\hbox{with $x_{s_1}<x_{s_2} < \cdots < x_{s_n}$.} 
$$
The condition that these lines form a consistent configuration
(i.e.\ that they do not intersect) can now be formulated as follows:
consider the free group generated by the $x_k$ subject to the
relations $x_k^2=1$. Then the configuration is consistent if the group
element   
\be\label{unit}
\Delta_1 \, \Delta_2 \, \cdots \, \Delta_r = 1
\ee
in the group.  To understand this, note that if $x_s$ appears to the
right of $x_r$ in (\ref{unit}), then the large-$x$ limit of  
the corresponding contour passing through $x_s$ lies counterclockwise
to the large-$x$ limit of the contour passing through $x_r$. Each
group element $x_k$ appears twice in (\ref{unit}), as each brane has
two branches.  Since the contours may not cross, we may never have a
sequence $x_s\ldots x_r \ldots x_s \ldots x_r$.  The allowed ordering
is: $x_s\ldots x_r\ldots x_r\ldots x_s$, $x_s,\ldots x_s,\ldots
x_r,\ldots x_r$, or the same with $x_s \leftrightarrow x_r$.  For $m$
finite this means that one will eventually find at least 
one $x_t$ adjacent to the second occurrence of $x_t$ in (\ref{unit}).
Using the relations, these can be removed and the process continued --
the result will be that all pairs are replaced by the identity, using
the relations. 

Any consistent configuration is essentially uniquely
characterised by the product of the $x_k$ (without reference to the
$\Delta_j$): given any product of $x_k$, we collect all increasing
elements in $\Delta_1$, the next sequence of increasing elements in
$\Delta_2$, etc. This is then also a consistent configuration and it
involves, if anything, fewer UV lines (i.e.\ fewer $\Delta_j$) than the
original configuration. So for example, for the sequence
$$
1 2 3 4 4 3 5 6 6 5 2 1 
$$
we take
$$
\Delta_{1}=\{1234\} \ , \ \
\Delta_{2}=\{4\} \ , \ \ 
\Delta_{3}=\{356\} \ , \ \ 
\Delta_{4}=\{6\} \ , \ \ 
\Delta_{5}=\{5\} \ , \ \ 
\Delta_{6}=\{2\} \ , \ \
\Delta_{7}=\{1\} \ .
$$
Thus we have reduced the problem to proving that (\ref{unit}) implies
that there are at least $r=m+1$ sets of increasing sequences. This can
be proved by induction on $m$.

The first step of the induction is obvious by inspection. Next, assume that
we have proven the statement for all sequences of pairs of $x_k$,
subject to the constraints above, with 
$k=1,\ldots,m\leq m_0-1$. Now consider a group element (\ref{unit})
involving the $m_0$ different $x_k$, $k=1,\ldots,m_0$. We concentrate
on the two elements $x_{m_0}$ that appear in (\ref{unit}). There are
two cases to consider.

If the two $x_{m_0}$ are adjacent in (\ref{unit}), it is clear that the
second $x_{m_0}$ makes up a $\Delta$ by itself. Thus if we remove
the two $x_{m_0}$ we reduce the number of $\Delta$s at least by one; 
the induction hypothesis then implies that the number of $\Delta_j$
must have been at least $m_0+1$.

In the other case, the two $x_{m_0}$ do not stand next to each
other. Then the group element in question must be of the form
\be\label{deco}
g_1 \, x_{m_0} \, h \, x_{m_0} \, g_1^{-1} \ . 
\ee
In order for this to be the identity it follows that $h$ must be the
identity (in the free group involving $r_h$ elements) and that $g_1$
and $g_1^{-1}$ can only involve the other $m_0-1-r_h$ elements. By the
induction hypothesis, $h$ involves at least $r_h+1$
$\Delta$s. Similarly, the product $g_1 g_1^{-1}$ involves at least
$m_0-r_h$ $\Delta$s. The first $x_{m_0}$ can always be adjoined to the
last $\Delta_j$ of $g_1$ (and it will guarantee that the first element
of $h$ will lie in a new $\Delta$), while the second $x_{m_0}$ can
always be adjoined to the last $\Delta_j$ of $h$ (and it will
guarantee that the first element of $g_1^{-1}$ will again lie in a new
$\Delta$). Thus it follows that the total number of $\Delta$s of
(\ref{deco}) is at least 
$$
(r_h+1) + (m_0-r_h) = m_0 + 1 \ ,
$$
proving our claim. \hspace{\fill}{\bf QED}.

\vskip .3cm

Next, we wish to show that all of the $L=0$ branes between $B$ and an
IR line can be generated from the decoupling branes. To see this we
first prove:

\vskip .2cm

\noindent {\bf Claim 2.} 
Suppose that there exists for each $\Delta_k$ an ascending sequence 
$(\Delta_k) = x_{s^k_1}\cdots x_{s^k_r}$, where 
$k_i\in\{1,\ldots, m\}$ such that  
\be\label{eq:masterrel}
	(\Delta_1)\cdots(\Delta_{m+1}) = 1 \ . 
\ee
[Note that it follows from Claim~1 that every consistent configuration
is of this form.] 
Then there is no proper subsequence
$S_1 = (\Delta_k)(\Delta_{k+1})\cdots(\Delta_l) = 1$ 
with $1 \leq k < l \leq m+1$: in other words,
the only such sequence is the case $k=1,l=m+1$.

\vskip .2 cm

\noindent {\bf Proof:}
If there were such a proper subsequence, then 
this sequence must be made from $m_1$ pairs of critical points
(as the algebra is freely  generated up to the relation $x_k^2 = 1$).
However, since by supposition (\ref{eq:masterrel}) is also satisfied,
this means that 
\be
S_2 = (\Delta_1)\cdots(\Delta_{k-1})(\Delta_{l+1})\cdots
(\Delta_{m+1}) = 1 \ . 
\ee
The sequence $S_2$ is then made up of $m_2$ pairs of critical points,
where by construction $m_1+m_2=m$. 

Now it follows from Claim~1 that $S_1$ must have at least $m_1 + 1$
$\Delta$'s, while $S_2$ must have at least $m_2 + 1$ $\Delta$'s, so
there must be at least $m_1 + m_2 + 2 = m+2$ $\Delta$s. But we know
there are only $m+1$ different $\Delta$s. It therefore follows that
such a proper subsequence is impossible. 
\nopagebreak
\hspace*{\fill} {\bf QED}.

\vskip .3cm

This is now enough to show that all $L=0$ branes between $B$ and an IR
line can be obtained by tachyon condensation from the decoupling
branes. Pick any two $\Delta_a$ and $\Delta_b$ in this range. 
We wish to obtain the brane which asymptotes to these two UV lines.
Such a brane can be built as follows.  We start with some critical
point in the sequence $\Delta_a$.  This will  also appear in some
other sequence $\Delta_{c_1}$.  There will be some critical point in 
$\Delta_{c_1}$ which also appears in a third sequence $\Delta_{c_2}$,
and so on until you reach $\Delta_b$. Thus there exists a brane
between $\Delta_a$ and $\Delta_{c_1}$, to which we can add a brane 
between $\Delta_{c_1}$ and $\Delta_{c_2}$ to get a brane between
$\Delta_a$ and $\Delta_{c_2}$. In turn we can add to this a brane
between $\Delta_{c_2}$ and $\Delta_{c_3}$ to get a brane between
$\Delta_a$ and $\Delta_{c_3}$, and so on until we obtain the desired
brane.  

There is {\it always} such a chain of pairs.  To see this, assume the
opposite, and start with the critical points in $\Delta_a$.  Now
consider the set $A_1$ of $\Delta$s which share a critical point with
$\Delta_a$.  Next consider the set $A_2$ of $\Delta$s which share a
critical point with $A_1$. Continue inductively.  The total number of
$\Delta$s (and of critical points) is finite, so this procedure must
terminate in a largest set $P$.  Because of the algebra and because of
Eq. (\ref{eq:masterrel}) and because $P$ does not include $\Delta_b$,
$P$ must be a sequence of $\Delta$s of type $S_1$. But we have already
shown that no such sequence exists.  Thus, a chain of pairs exists
such that one can build a bound state of decoupling branes that
asymptotes to any adjacent pair of UV lines between $B$ and the IR
line.

\section{The configuration of decoupling branes for $d$ odd}

We will consider the case $\lambda \in \IR$. Then it follows from 
eq. (\ref{eq:critpoints}) that all the critical points lie on a 
circle of radius $R(\lambda)$ in the $x$-plane.  

\vskip .2cm
\noindent{\bf Claim 3.}
Consider the first and last critical point, that is, $x_0$ and $x_u$
in \S4.2, that have the same value of the imaginary part of $W$. There
is a contour with $\Imm\ W = {\rm const}$ that runs between them  
{\it inside} the circle of radius $R$. 

\vskip .2cm
\noindent{\bf Proof:}
As one moves counterclockwise along the circle of radius
$R$ from $x_0$ to $x_u$, $\Imm\ W$ increases first and then
decreases; in particular, $\Imm\ W$ is bigger than the value at
$x_0$ (or $x_u$) at any point $x'$ on the circle between $x_0$
and $x_u$. Consider the straight line from the origin to any point
$x'$. At the origin $\Imm\ W=0$ and hence by the intermediate 
value theorem, there is a point inside the 
circle of radius $R$ where $\Imm\ W$ has the same value as
at $x_0$ (see figure~8). Now, scan $x'$ along the circle from 
$x_0$ to $x_u$ and we thus sweep out the contour of constant 
$\Imm\ W$ inside the circle. \hspace{\fill} {\bf QED}. 

\begin{figure}[htb]
\centering
\epsfig{file=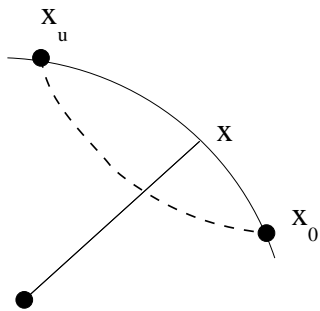,height=4cm}
\parbox{5.5in}{\caption{The line from the origin to $x'$.  The dashed
line is t`he contour of constant $\Imm\ W$ running from $x_0$ to
$x_u$.}} 
\end{figure}

\vskip .2cm

Note that the same argument shows that there is such a contour, which
we denote $a_k$, running between $x_k$ and $x_{u-k}$ for any 
$k < \frac{u}{2}$. Furthermore,  these contours cannot cross, so they
are nested inside each other. 

Now, since $\Ree\ W$ increases between $x_{u-k}$ and $x_k$, the
contours we have drawn must be one of the two branches of the D-brane
emanating from $x_{u-k}$. The other branch, which we will call
$b_{u-k}$, must asymptote to a UV line. For the D-brane emanating from
$x_k$, there must be two asymptoting to UV lines, and we will denote
by $b_k$ the contour running from $x_k$ to the UV line closest to the
real axis. 

Define the contour $c_k = b_k \cup b_{u-k}\cup a_k$ for 
$k \leq \frac{u}{2}$.  If $\frac{u}{2}\in \IZ$, 
define $c_{\frac{u}{2}}$ to be the brane emanating from this critical
point as well. Now since the UV lines may not cross, and the contours
$a_k$ are nested inside each other inside the circle of radius $R$,
$c_k$ must be contained inside $c_{k'}$ if $k' > k$. 

By Claim~1 of appendix~A, all of the UV lines between the real 
axis and $B$ must have at least one decoupling brane asymptoting
towards it. Since contours with different values of $\Imm\ W$ cannot
cross, and the contours $c_k$ are nested inside each other, then 
$b_u$ must asymptote to the UV line closest in a clockwise direction 
to $B$, and $b_1$ must asymptote to the real axis.
Similarly, $b_{u-1}$ must asymptote to the second closest UV line in a
clockwise direction from $B$, and $b_2$ must asymptote to the first UV
line in a counterclockwise direction from the real axis.   

Finally, call the $d_k$ the branch of the brane emanating from 
$x_{k <\frac{u}{2}}$  that is not $b_k$.  $d_k$ must asymptote to the
UV line immediately counterclockwise  from $b_k$. This follows from
two facts. First, as discussed in \S4.2, two contours  of constant
$\Imm\ W$ which asymptote to the same UV line are ordered in  a
counterclockwise direction, such that the contour with largest 
$\Imm\ W$ is counterclockwise to the other contour as
$|x|\to\infty$. Secondly, contours of constant $\Imm\ W$ with differing
values of $\Imm\ W$ cannot cross. This leaves only one option, namely
the one discussed.


\begin{thebibliography}{99}

\bibitem{Adams:2001sv}
  A.~Adams, J.~Polchinski and E.~Silverstein,
  {\it Don't panic! Closed string tachyons in ALE space-times},
  JHEP {\bf 0110}, 029 (2001)
  {\tt  [arXiv:hep-th/0108075]}.
  
\bibitem{Martinec:2002wg}
  E.J.~Martinec and G.W.~Moore,
  {\it On decay of K-theory},
  {\tt arXiv:hep-th/0212059}.
  
\bibitem{Minwalla:2003hj}
  S.~Minwalla and T.~Takayanagi,
  {\it Evolution of D-branes under closed string tachyon condensation},
  JHEP {\bf 0309}, 011 (2003)
  {\tt  [arXiv:hep-th/0307248]}.
 
\bibitem{Moore:2004yt}
  G.W.~Moore and A.~Parnachev,
  {\it Localized tachyons and the quantum McKay correspondence},
  JHEP {\bf 0411}, 086 (2004)
  {\tt [arXiv:hep-th/0403016]}.
  
\bibitem{Moore:2005wp}
  G.W.~Moore and A.~Parnachev,
  {\it Profiling the brane drain in a nonsupersymmetric orbifold},
  JHEP {\bf 0601}, 024 (2006)
  {\tt [arXiv:hep-th/0507190]}.

\bibitem{Adams:2005rb}
  A.~Adams, X.~Liu, J.~McGreevy, A.~Saltman and E.~Silverstein,
  {\it Things fall apart: Topology change from winding tachyons},
  JHEP {\bf 0510}, 033 (2005)
  {\tt [arXiv:hep-th/0502021]}.

\bibitem{Ghoshal:1993tm}
  S.~Ghoshal and A.B.~Zamolodchikov,
  {\it Boundary S matrix and boundary state in two-dimensional 
	integrable quantum field theory},
  Int.\ J.\ Mod.\ Phys.\ A {\bf 9}, 3841 (1994)
  [Erratum-ibid.\ A {\bf 9}, 4353 (1994)]
  {\tt [arXiv:hep-th/9306002]}.

\bibitem{LeClair:1995uf}
  A.~LeClair, G.~Mussardo, H.~Saleur and S.~Skorik,
  {\it Boundary energy and boundary states in integrable quantum field
	theories},  
  Nucl.\ Phys.\ B {\bf 453}, 581 (1995)
  {\tt [arXiv:hep-th/9503227]}.

\bibitem{Chatterjee:1995be}
  R.~Chatterjee,
  {\it Exact partition function and boundary state of 2-D massive 
	Ising field theory with boundary magnetic field},
  Nucl.\ Phys.\ B {\bf 468}, 439 (1996)
  {\tt [arXiv: hep-th/9509071]}.
  
\bibitem{Dorey:1997yg}
  P.~Dorey, A.~Pocklington, R.~Tateo and G.M.T.~Watts,
  {\it TBA and TCSA with boundaries and excited states},
  Nucl.\ Phys.\ B {\bf 525}, 641 (1998)
  {\tt [arXiv:hep-th/9712197]}.
  
\bibitem{Dorey:1999cj}
  P.~Dorey, I.~Runkel, R.~Tateo and G.M.T.~Watts,
  {\it g-function flow in perturbed boundary conformal field theories},
  Nucl.\ Phys.\ B {\bf 578}, 85 (2000)
  {\tt [arXiv:hep-th/9909216]}.
  
\bibitem{Dorey:2004xk}
  P.~Dorey, D.~Fioravanti, C.~Rim and R.~Tateo,
  {\it Integrable quantum field theory with boundaries: 
	The exact g-function},
  Nucl.\ Phys.\ B {\bf 696}, 445 (2004)
  {\tt [arXiv:hep-th/ 0404014]}.

\bibitem{Green:2006ku}
   D.~Green,
   {\it Nothing for branes},
   {\tt arXiv:hep-th/0611003}.

\bibitem{Fredenhagen:2006dn}
S.~Fredenhagen, M.R.~Gaberdiel and C.A.~Keller,
  {\it Bulk induced boundary perturbations},
  J.\ Phys.\  A {\bf 40}, F17 (2007) 
  {\tt [arXiv:hep-th/0609034]}.

\bibitem{Freedman:2005wx}
  D.Z.~Freedman, M.~Headrick and A.~Lawrence,
  {\it On closed string tachyon dynamics},
  Phys.\ Rev.\ D {\bf 73}, 066015 (2006)
  {\tt [arXiv:hep-th/0510126]}.
  
\bibitem{Graham:2006gc}
  K.~Graham, A.~Konechny and J.~Teschner,
  {\it On the time-dependent description for the decay of unstable
  D-branes},
  {\tt arXiv:hep-th/0608003}.

\bibitem{McGreevy:2005ci}
  J.~McGreevy and E.~Silverstein,
  {\it The tachyon at the end of the universe},
  JHEP {\bf 0508}, 090 (2005)
  {\tt [arXiv:hep-th/0506130]}.

\bibitem{Cardy:1989ir}
   J.L.~Cardy,
   {\it Boundary conditions, fusion rules and the Verlinde formula},
   Nucl.\ Phys.\ B {\bf 324}, 581 (1989).

\bibitem{Cvetic:1989qv}
   M.~Cvetic and D.~Kutasov,
   {\it Topology change in string theory},
   Phys.\ Lett.\ B {\bf 240}, 61 (1990).
  
\bibitem{HIV} 
   K.~Hori, A.~Iqbal and C.~Vafa,
   {\it D-branes and mirror symmetry},
   {\tt arXiv:hep-th/0005247}.

\bibitem{Affleck:1991tk}
   I.~Affleck and A.W.W.~Ludwig,
   {\it Universal noninteger 'ground state degeneracy' in critical
   quantum systems},
   Phys.\ Rev.\ Lett.\  {\bf 67}, 161 (1991).

\bibitem{Recknagel:1998sb}
   A.~Recknagel and V.~Schomerus, 
   {\it D-branes in {Gepner} models},
   Nucl.\ Phys. B {\bf 531}, 185 (1998) 
   {\tt [arXiv:hep-th/9712186]}.

\bibitem{Zamolodchikov:1986db}
   A.B.~Zamolodchikov,
   {\it Conformal symmetry and multicritical points in
   two-di\-men\-sio\-nal quantum field theory},
   Sov.\ J.\ Nucl.\ Phys.\  {\bf 44}, 529 (1986)
   [Yad.\ Fiz.\  {\bf 44}, 821  (1986)].

\bibitem{Kastor:1988ef}
   D.A.~Kastor, E.J.~Martinec and S.H.~Shenker,
   {\it RG flow in N=1 discrete series},
   Nucl.\ Phys.\ B {\bf 316}, 590 (1989).
  
\bibitem{Vafa:1988uu}
   C.~Vafa and N.P.~Warner,
   {\it Catastrophes and the classification of conformal theories},
   Phys.\ Lett.\ B {\bf 218}, 51 (1989).

\bibitem{Martinec:1988zu}
   E.J.~Martinec,
   {\it Algebraic geometry and effective Lagrangians},
   Phys.\ Lett.\ B {\bf 217}, 431 (1989).
  
\bibitem{Hori:2000kt}
  K.~Hori and C.~Vafa,
  {\it Mirror symmetry},
  {\tt arXiv:hep-th/0002222}.

\bibitem{Vafa:2001ra}
  C.~Vafa,
  {\it Mirror symmetry and closed string tachyon condensation},
  {\tt arXiv:hep-th/ 0111051}.
  
\bibitem{Ooguri:1996ck}
H.~Ooguri, Y.~Oz and Z.~Yin,
   {\it D-branes on Calabi-Yau spaces and their mirrors},
   Nucl.\ Phys.\ B {\bf 477}, 407 (1996) 
   {\tt [arXiv:hep-th/9606112]}.

\bibitem{Fredenhagen:2001nc}
S.~Fredenhagen and V.~Schomerus,
{\it Brane dynamics in CFT backgrounds},
{\tt arXiv: hep-th/0104043}.

\bibitem{Maldacena:2001ky}
J.M.~Maldacena, G.W.~Moore and N.~Seiberg,
   {\it Geometrical interpretation of D-branes in gauged WZW models},
   JHEP {\bf 0107}, 046 (2001) 
   {\tt [arXiv:hep-th/0105038]}.

\bibitem{aelunpub}
A.~Lawrence, unpublished calculation, described in
\cite{Green:2006ku}. 

\bibitem{Hori:2000ic}
  K.~Hori,
  {\it Linear models of supersymmetric D-branes,}
  {\tt arXiv:hep-th/0012179}.

\bibitem{Harvey:1999gq}
J.A.~Harvey, S.~Kachru, G.W.~Moore and E.~Silverstein,
   {\it Tension is dimension},
   JHEP {\bf 0003}, 001 (2000) 
   {\tt [arXiv:hep-th/9909072]}.

\bibitem{Friedan:2003yc}
D.~Friedan and A.~Konechny,
   {\it On the boundary entropy of one-dimensional quantum systems at low
   temperature},
   Phys.\ Rev.\ Lett.\  {\bf 93}, 030402 (2004)
   {\tt [arXiv:hep-th/0312197]}.

\bibitem{Brunner:2003zm}
I.~Brunner and K.~Hori,
  {\it Orientifolds and mirror symmetry},
  JHEP {\bf 0411}, 005 (2004) 
  {\tt [arXiv:hep-th/0303135]}.

\bibitem{BR} I.~Brunner and D.~Roggenkamp, in preparation.
  
\bibitem{Kont:unpub}
M.~Kontsevich, unpublished work.

\bibitem{Kapustin:2002bi}
  A.~Kapustin and Y.~Li,
  {\it D-branes in Landau-Ginzburg models and algebraic geometry},
  JHEP {\bf 0312}, 005 (2003)
  {\tt [arXiv:hep-th/0210296]}.

\bibitem{Brunner:2003dc}
I.~Brunner, M.~Herbst, W.~Lerche and B.~Scheuner, 
  {\it Landau-Ginzburg realization of open string TFT},
JHEP {\bf 0611}, 043 (2006)  {\tt [arXiv:hep-th/0305133]}.

\bibitem{Herbst:2004zm}
M.~Herbst, C.I.~Lazaroiu and W.~Lerche,
{\it D-brane effective action and tachyon condensation in topological
minimal models},
JHEP {\bf 0503}, 078 (2005) 
{\tt [arXiv:hep-th/0405138]}.

\bibitem{Bourdeau:1991uu}
  M.~Bourdeau, E.J.~Mlawer, H.~Riggs and H.J.~Schnitzer,
  {\it Topological Landau-Ginzburg matter from Sp(N)-K fusion rings},
  Mod.\ Phys.\ Lett.\ A {\bf 7}, 689 (1992)
  {\tt [arXiv: hep-th/9111020]}.

\bibitem{Recknagel:2000ri}
  A.~Recknagel, D.~Roggenkamp and V.~Schomerus,
  {\it On relevant boundary perturbations of unitary minimal models},
  Nucl.\ Phys.\ B {\bf 588}, 552 (2000)
  {\tt [arXiv:hep-th/0003110]}.

\bibitem{Affleck:1990by}
  I.~Affleck and A.W.W.~Ludwig,
  {\it The Kondo effect, conformal field theory and fusion rules},
  Nucl.\ Phys.\ B {\bf 352}, 849 (1991).

\bibitem{Fredenhagen:2003xf}
  S.~Fredenhagen,
  {\it Organizing boundary RG flows},
  Nucl.\ Phys.\ B {\bf 660}, 436 (2003)
  {\tt [arXiv:hep-th/0301229]}.


\end{thebibliography}
\end{document}